\documentclass[twocolumn]{aastex631}
\newcommand\mstar{M_{*}}
\newcommand\msun{M_{\odot}}
\newcommand\mhi{M_{HI}}
\newcommand\LG{L_{g}}
\newcommand\lsun{L_{\odot}}
\newcommand\vsys{V_{sys}}
\newcommand\wfty{W_{50}}
\newcommand\wftyc{W_{50,c}}
\newcommand\wftyct{W_{50,c,t}}
\newcommand{\kms}{{\ensuremath{\mathrm{km\,s^{-1}}}}}
\newcommand\mhilim{M^{lim}_{HI}}
\newcommand\mbary{M_{bary}}
\defcitealias{2020Karunakaran}{K20}
\defcitealias{2023Zaritsky}{Z23}
\defcitealias{2021Zaritsky}{Z21}
\defcitealias{2022Zaritsky}{Z22}
\begin{document}

\title{Systematically Measuring Ultra-Diffuse Galaxies. VII. The \ion{H}{1} Survey Overview}

\correspondingauthor{Ananthan Karunakaran}
\email{ananthan.karunakaran@utoronto.ca}

\author[0000-0001-8855-3635]{Ananthan Karunakaran}
\affiliation{Department of Astronomy \& Astrophysics, University of Toronto, Toronto, ON M5S 3H4, Canada}
\affiliation{Dunlap Institute for Astronomy and Astrophysics, University of Toronto, Toronto ON, M5S 3H4, Canada}

\author{Khadeejah Motiwala}
\affiliation{Department of Physics, Engineering Physics and Astronomy, Queen’s University, Kingston, ON K7L 3N6, Canada}

\author[0000-0002-0956-7949]{Kristine Spekkens}
\affiliation{Department of Physics, Engineering Physics and Astronomy, Queen’s University, Kingston, ON K7L 3N6, Canada}

\author[0000-0002-5177-727X]{Dennis Zaritsky}
\affiliation{Steward Observatory, University of Arizona, 933 North Cherry Avenue, Tucson, AZ 85721-0065, USA}

\author[0000-0001-7618-8212]{Richard L. Donnerstein}
\affiliation{Steward Observatory, University of Arizona, 933 North Cherry Avenue,  Tucson, AZ 85721-0065, USA}

\author[0000-0002-4928-4003]{Arjun Dey}
\affiliation{NSF's National Optical-Infrared Astronomy Research Laboratory, 950 N. Cherry Ave., Tucson, AZ 85719, USA}

\begin{abstract}

We present the results from the neutral hydrogen (\ion{H}{1}) follow-up survey of 378 optically-detected UDG candidates from the Systematically Measuring Ultra-Diffuse Galaxies (SMUDGes) survey using the Robert C.\ Byrd Green Bank Telescope (GBT).\ We detect \ion{H}{1} in 110 targets and determine 37 to be UDGs and 73 to be low surface brightness (LSB) dwarfs based on their effective radii and central surface brightnesses.\ In line with previous studies, we find that: i) our \ion{H}{1} detections have on average bluer $g-r$ colors and more irregular morphologies than our \ion{H}{1} non-detections, ii) our \ion{H}{1} detections populate the tail end of the star-forming main sequence from the ALFALFA catalog with marginally lower specific star formation rates, and iii) \ion{H}{1} detections are mostly in relatively isolated (i.e.\ field) environments, while most non-detections have at least one nearby neighbor in projection.\ We find that the \ion{H}{1} mass to stellar mass ratios (i.e.\ gas richnesses) scale with the physical sizes for UDGs and LSB dwarfs alike, suggesting that mechanisms other than bursty star formation feedback may be at play for UDGs.\ However, we find a stronger trend between gas richnesses and physical sizes if we define UDGs using their effective surface brightness instead of their central surface brightness.\ We are in the process of using this unprecedented sample of UDG candidates to carry out detailed follow-up studies (i.e.\ star-formation and environmental analysis, comparisons to simulations) and are obtaining resolved \ion{H}{1} observations for several of them.\

\end{abstract}


\section{Introduction} \label{sec:intro}
The faintest galaxies play a crucial role in the hierarchical nature of galaxy formation and evolution with the $\Lambda$CDM paradigm and yet pose significant challenges to it \citep[e.g.][]{2017Bullock,2022Sales}.\ Throughout the last decade, there has been significant work on studying some of the faintest galaxies in the local Universe.\ Much of this work has been fueled by the advent of both dedicated surveys using bespoke facilities (e.g.\ Dragonfly Telephoto Array, \citealt{DragonflyArrayOriginal}) and deep, wide-field surveys with larger observatories (e.g.\ Legacy Surveys, \citealt{Dey2019desi} and Dark Energy Survey, \citealt{2018DES}).\ Arguably one of the most interesting results from these studies was the detection of a large number of extended $(R_{\mathrm{eff}} \geq 1.5 \,\mathrm{kpc})$\footnote{We use $r_{\mathrm{eff}}$ for the angular effective radius and $R_{\mathrm{eff}}$ for the physical effective radius throughout this paper.}, low surface brightness $(\mu_{0,g}\geq24\,\mathrm{mag/arcsec^2})$ (LSB) galaxies in the Coma Cluster \citep{2015vandokkum}.\ These galaxies, commonly referred to as ultra-diffuse galaxies (UDGs), are similar to the large LSB galaxies detected in studies using photographic plates and early on in the introduction of CCDs \citep{SandageUDG,1988Impey,1991Bothun,1997ONeil} and have been suggested to be an extension of the dwarf elliptical/spheroidal population \citep{Conselice}.\

Since their (re-)discovery, there have been hundreds of studies focused on UDGs.\ Thousands of UDG candidates have been detected in clusters \citep[e.g.,][]{2015koda,2016Yagi,SMUDGes,2015Mihos,2016Beasleya,2017Shi,2017vanderBurg,2019Janssens,2019ManceraClusters,2020Lee,lim2020,2022LaMarca,2022Venhola}, groups \citep[e.g.][]{2017RomanTrujillo,muller2018,2020Somalwar,2021Gannon,2023Goto,2023Karunakaran,2024Jones}, and lower density field-like environments \citep[e.g.,][]{2016MartinezDelgado,2017Bellazzini,2017Leisman,2017Trujillo,2018bennet,2019prole,2019Roman,2020SPLUS,2021Tanoglidis,2024Fielder,2024Montes}.\ As seen for more massive, higher surface brightness galaxies, UDGs and UDG candidates follow the morphology-density relation \citep{Dressler}, with those those in less dense environments having younger (i.e.\ bluer) stellar populations, more irregular morphologies, and substantial neutral hydrogen (\ion{H}{1}) reservoirs \citep[e.g.][]{2017Leisman,2019prole,2021Kadowaki}, while those in dense environments have older (i.e.\ redder) stellar populations and smoother morphologies \citep[e.g.][]{lim2020}.\

Across these environments, there exists a large diversity in the physical properties of UDGs, similar to that seen in the high surface brightness galaxy population.\ Most UDGs seem to be embedded in dwarf galaxy-mass dark matter halos \citep[e.g.,][]{2016Beasleyb,2018Amorisco,2019Chilingarian,2019Prolehalomass}, although some UDGs exhibit extreme properties that pose challenges to proposed galaxy formation mechanisms, such as high dark matter fractions \citep{2016vandokkum,2016Beasleya}, dark matter deficiencies (\citealt{2018vandokkum,2019vandokkum}; although see \citealt{2019Trujillo}), and offsets from established galaxy scaling relations such as the baryonic Tully-Fisher relation (\citealt{2019HUDsBTFR,2020ManceraPina}; although see \citealt{2020Karunakaran}).\ 

Understanding how such extended systems can form within these dark matter halos has been studied in great detail and these mechanisms typically invoke either internal or external processes to form UDGs.\ In the latter scenario, normal dwarf galaxies may be physically transformed to have more extended sizes through tidal interactions by a massive companion \citep[e.g.][]{2018bennet,2020Sales,2021Jones,2024Fielder} or are formed as UDGs in the field and subsequently quenched through tidal and ram-pressure stripping in more dense environments \citep[e.g.][]{2015YozinBekki,2019Liao,2019Jiang,2019Carleton,2020Tremmel,2023Benavides}.\ There are several proposed formation mechanisms for UDGs in lower-density environments.\ Isolated UDGs may form through internal processes such as bursty star formation that redistributes baryonic matter to larger radii \citep[e.g.][]{2017dicintio} or form within high-spin dark matter halos which dictate their stellar distributions \citep[e.g.][]{2016AmoriscoLoeb}.\ It has also been suggested that external processes can form isolated UDGs such as dwarf-dwarf mergers at early times in the Universe \citep[e.g.][]{2021Wright}.\ Two crucial components necessary to begin to unequivocally constrain different UDG formation mechanisms are a uniformly selected sample of UDG candidates and distance measurements to confirm these candidates as true UDGs and determine their physical properties.\ 

The Systematically Measuring Ultra-Diffuse Galaxies \citep[SMUDGes,][hereafter Z23]{2023Zaritsky} survey began as a focused endeavor to detect UDG candidates around the Coma Cluster and its environment \citep{SMUDGes} using optical imaging from the pre-imaging surveys for the Dark Energy Survey Instrument (DESI) survey \citep[now commonly referred to as the Legacy Surveys,][]{Dey2019desi}.\ The complete SMUDGes survey has detected over 7000 UDG candidates based on their angular sizes $(r_{\mathrm{eff}}\geq5.3'')$ and central surface brightness $(\mu_{0,g}\geq24\,\mathrm{mag\,arcsec^{-2}})$ across 20,000 square degrees of the Legacy Surveys footprint.\ These UDG candidates span a wide range of environments from dense clusters to more sparse regions.\ 

Several UDG studies have focused on these more dense environments and can take advantage of distance-by-association to confirm any detected UDG candidates as true UDGs.\ Due to their intrinsic faintness, obtaining distance through optical spectroscopy is difficult and there are on the order of $\sim100$ UDGs with optical spectroscopic distance measurements (e.g.\ see \citealt{2021Kadowaki} and \citealt{2024Gannon} for compilations).\ It is well-established that UDGs in isolation follow the similar trends as normal dwarfs galaxies: they are typically blue, star-forming, and gas-rich \citep[e.g.][]{2017Papastergis,2017Leisman,HCGUDGs,2022Poulain} and simulations \citep[e.g.][]{2021Wright, 2023Benavides}.\

We initiated a pilot \ion{H}{1} follow-up survey of 70 UDG candidates selected from the SMUDGes sample in the 270 square degree region around the Coma Cluster \citep[][hereafter K20]{2020Karunakaran}.\ Using the Robert C.\ Byrd Green Bank Telescope, we detected the \ion{H}{1} reservoirs of 18 UDG candidates and placed stringent \ion{H}{1} upper limits on the remainder of the sample.\ Using the kinematic distances of these systems, we determined that nine of these systems have physical sizes that satisfy the aforementioned UDG criteria, (i.e.\ $R_{\mathrm{eff}} > 1.5\,\mathrm{kpc}$).\ These positive first results were useful for establishing a much larger \ion{H}{1} follow-up survey of SMUDGes UDG candidates using the GBT.\ 

In this paper, we present the results from our \ion{H}{1} survey along the lines of sight to 378 SMUDGes UDG candidates.\ The goals of this survey are to: i) obtain distance estimates to a significant sample of gas-rich SMUDGes candidates across a variety of environments, ii) estimate their \ion{H}{1} properties, iii) make comparisons of the baryonic properties of our sample to predictions from simulations, and iv) build up a sample of targets for more detailed \ion{H}{1} follow-up imaging.\ 

The structure of this paper is as follows.\ In Section \ref{sec:sample}, we describe our \ion{H}{1} survey design and sample selection and move on to present the observations and data reduction in Section \ref{sec:Obsanddata}.\ In Section \ref{sec:results}, we present the properties of our \ion{H}{1} detections and non-detections as well as preliminary results on their star-forming and environmental properties.\ In Section \ref{sec:Discussion}, we present a more detailed comparison between our \ion{H}{1} detections and non-detections before moving on to discuss UDG formation mechanisms, and finally discussing the effects of UDG definitions.\ We conclude and outline future work in Section \ref{sec:Conclusion}.\ Throughout this work we use $H_0 = 70 \kms \mathrm{Mpc^{-1}}$, $\Omega_\Lambda = 0.7$, and $ \Omega_m = 0.3.$

\section{Survey and Sample Description} \label{sec:sample}
We designed our survey and compiled our sample to ensure that we covered a wide range of environments.\ Our survey consists of five observing ``Campaigns'' corresponding to different observing semesters, each observing a sample of targets selected from preliminary SMUDGes pipeline catalogs.\ Targets were selected from preliminary candidate lists because the \ion{H}{1} follow-up samples were typically compiled and observed prior to the SMUDGes catalog being finalized and published.\ As a result, photometric properties used to compile the samples in our observing campaigns are slightly different from the final SMUDGes values \citepalias[see][for more details]{2021Zaritsky,2022Zaritsky} adopted in this work.\ We describe each of our campaigns below and summarize their key details in Table \ref{tab:surveysummary}.\

The primary selection criterion for the first three campaigns in our survey was the candidates' apparent $g-$band magnitude, $m_g$.\ All targets were selected to have $m_g\lesssim19.5\,\mathrm{mag}$ which enables a search for an \ion{H}{1} reservoir to sufficiently meaningful depths, i.e.\ to reach \ion{H}{1} mass-$g-$band luminosity ratios $\mhi/\LG = 1\,\msun/\lsun$.\ In our pilot survey/Campaign 1 \citepalias{2020Karunakaran}, we observed a total of 70 targets mostly from the region centered on the Coma Cluster which largely stem from the pilot SMUDGes catalog \citep{SMUDGes}.\ In Campaign 2, we selected 37 additional candidates from the region around the Coma Cluster and 26 candidates detected in the Stripe 82 region \citep[][]{2021Zaritsky} for a total of 63 targets.\ For Campaign 3, we selected our sample from an expanded search for candidates in an annulus covering between $10-20$ deg centered on the Coma Cluster, avoiding the Virgo Cluster region (for which crowding is problematic for the relatively wide GBT beam, and where few gas-rich UDGs are expected to exist).\ In addition to the 42 targets from this expanded Coma region search, we selected 16 targets from the Stripe82 sample for a total of 58 targets.\

Learning from the preliminary results of Campaigns~1-3 and in an effort to increase the \ion{H}{1} detection rates in Campaigns~4-5, we applied an additional criterion to select our targets: candidates must show UV emission in archival GALEX imaging.\ In Campaign 4, we selected 92 candidates from a strip across the DECaLS footprint stretching between $0\lesssim\mathrm{RA\,(deg)}\lesssim150$ and $23\lesssim\mathrm{Dec\,(deg)}\lesssim35$.\ We took a similar approach in selecting the sample for Campaign 5, selecting 95 targets from a strip stretching between $120\lesssim\mathrm{RA\,(deg)}\lesssim360$ and $2\lesssim\mathrm{Dec\,(deg)}\lesssim15$, again avoiding the Virgo Cluster.\ Additionally, for Campaign 5, we imposed a marginally brighter apparent magnitude cut of $m_g <19\,\mathrm{mag}$ and surface brightness cut of $\mu_{0,g} \gtrsim23.5\,\mathrm{mag/arcsec^2}$.\ While this expanded surface brightness selection will include many non-UDGs, it provides better overlap with other \ion{H}{1} samples of UDGs.\ For example, the \ion{H}{1}-bearing Ultra-Diffuse sources \citep[HUDS,][]{2017Leisman} selected from the ALFALFA sample where the restricted sample HUDS-R ($r_{(g,\mathrm{eff})} > 1.5\, \mathrm{kpc},\, \mu_{0,g} \gtrsim24\,\mathrm{mag/arcsec^2}$) matches well with our samples from Campaigns 1-4 and the broad sample HUDS-B ($r_{(r,\mathrm{eff})} > 1.5\,\mathrm{kpc},\, \langle\mu(r,r_{\mathrm{eff}})\rangle > 24\,\mathrm{mag/arcsec^2}$) with our sample from Campaign 5.

\begin{figure*}[htb!]
\includegraphics[width=18cm]{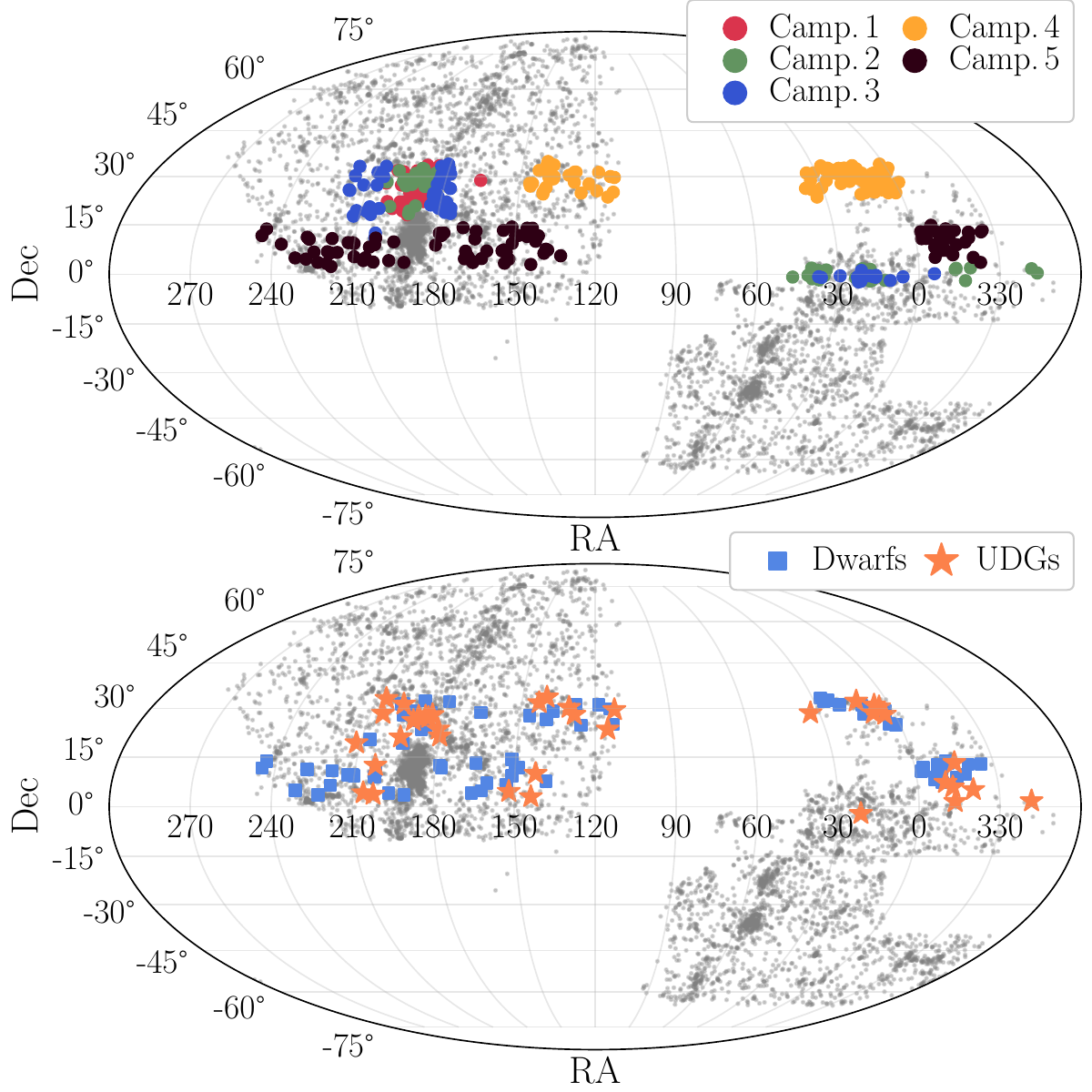}
\caption{Sky distribution of our UDG candidate \ion{H}{1} follow-up sample.\ In the top panel, we show all of our targets colored by the campaign in which they were observed.\ In the bottom panel, we show the distribution of our \ion{H}{1} detections: orange stars show the confirmed UDGs and the blue squares show the dwarf galaxies.\ In both panels we show the complete SMUDGes sample \citepalias{2023Zaritsky} as gray circles.\ }
\label{fig:skydist}
\end{figure*}

In total, our survey target sample consists of 378 SMUDGes candidates across five observing campaigns.\ We show the sky distribution of our survey sample in the top panel of Figure \ref{fig:skydist} with the symbol colors denoting their campaign.\ For reference, we also show the complete SMUDGes sample \citepalias{2023Zaritsky} as the small gray circles.\ We summarize key details (e.g.\ sample sizes, detection numbers, etc.) of our survey in Table \ref{tab:surveysummary}.

\begin{figure*}[htb!]
\includegraphics[width=8.5cm]{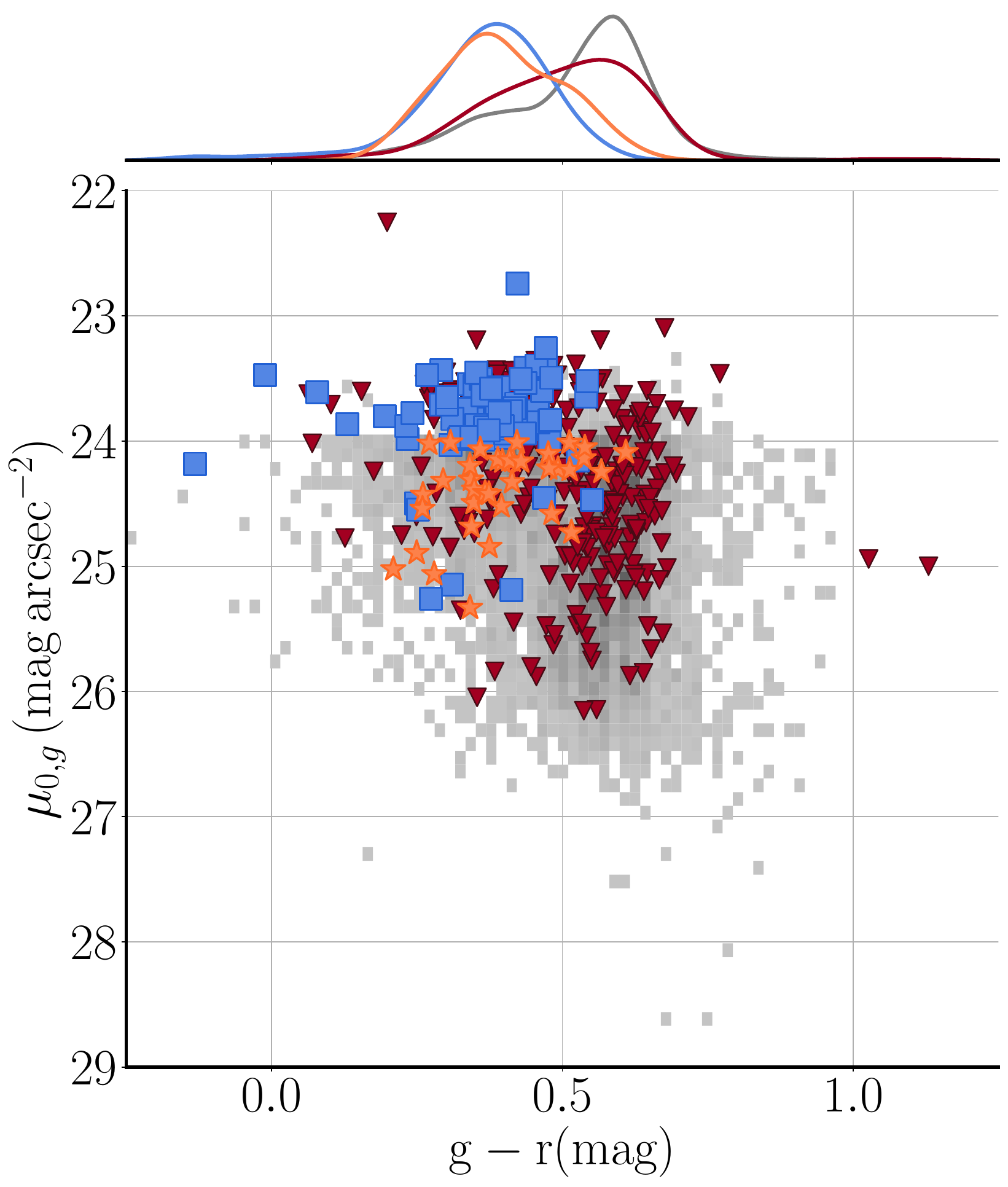}
\includegraphics[width=9.075cm]{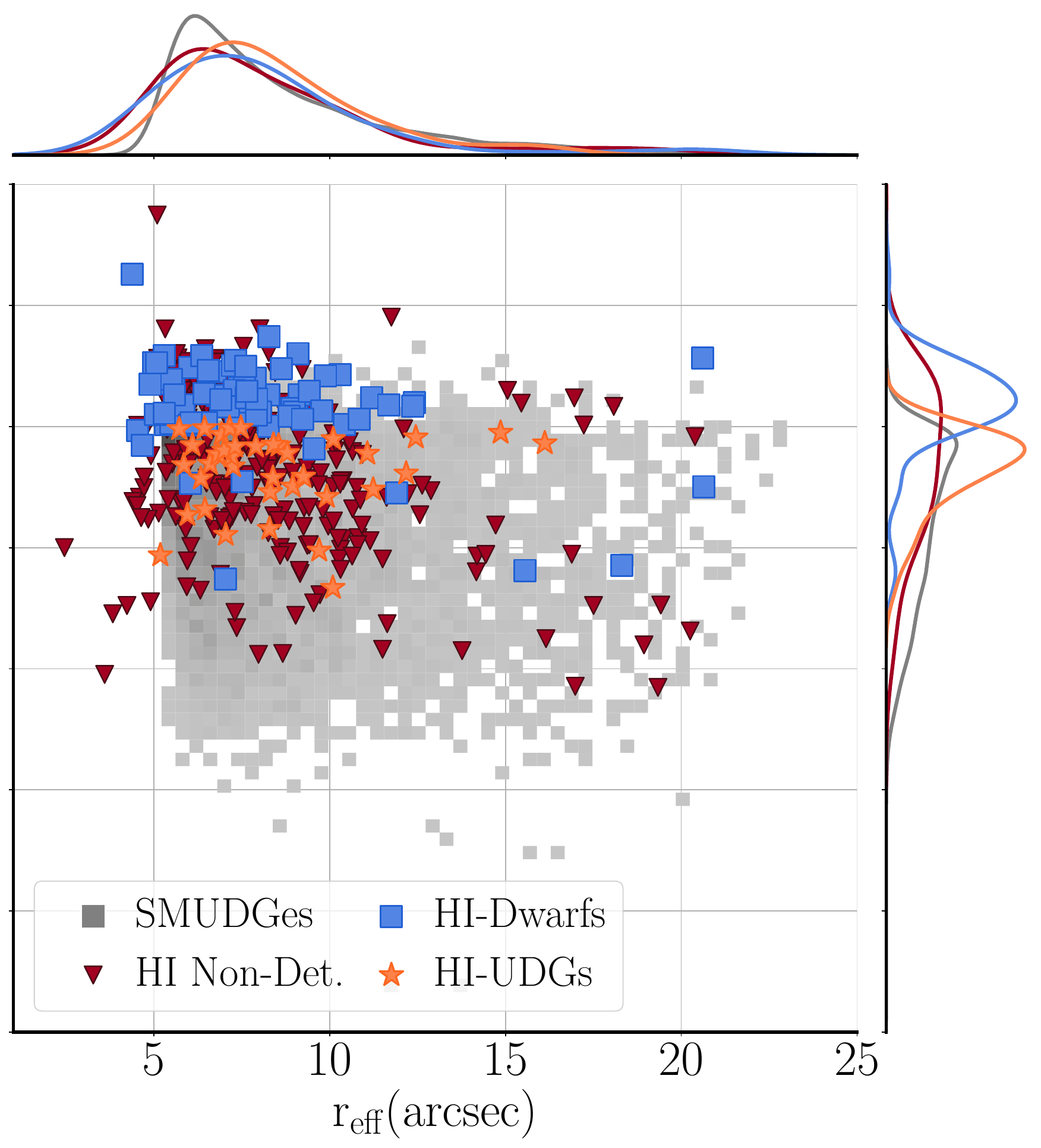}
\caption{Comparison of the apparent properties for our sample: central $g$-band surface brightness as a function of $g-r$ color (left panel) and apparent size (${r_{\mathrm{eff}}}$; right panel).\ Confirmed UDGs and dwarf galaxies are shown as orange stars and blue squares, respectively.\ \ion{H}{1} non-detections are shown as red inverted triangles and the grayscale heat map shows the complete SMUDGes sample.\ The marginalized distributions of each property are shown along the opposite axes.}
\label{fig:optpropcomp}
\end{figure*}

Given these differences and the expanded surface brightness criteria in Campaign 5, only $\sim 40\%$ (166/378) of the survey sample is present in the final SMUDGes catalog, although all were selected using the SMUDGes pipeline.\ They simply miss on one or more of the SMUDGes catalog criteria \citepalias{2023Zaritsky}.\ For the remaining sources, we are presenting the photometric properties, derived in the same manner as the SMUDGes catalog, for the first time.\ All photometric values in figures and tables are bias-corrected \citepalias[see][for details]{2023Zaritsky}.\ The photometric properties of the remaining $60\%$ are broadly commensurate with the final SMUDGes sample, aside from the fact that the sources have brighter central surface brightness. This can be seen in Figure \ref{fig:optpropcomp}, where the colored symbols show our survey sample and the grayscale heatmap shows the complete SMUDGes sample.\

\begin{deluxetable*}{ccCcCCC}[htb!]
\caption{Summary of SMUDGes in \ion{H}{1} Survey}
\label{tab:surveysummary}
\tablehead{
\colhead{Campaign} & \colhead{GBT Program ID} & \colhead{$N_{targets}$} & \colhead{Total Obs.\ Time} & \colhead{$N_{Det.}$} & \colhead{$N_{Dwarf}$} & \colhead{$N_{UDG}$}\\
\colhead{(1)} & \colhead{(2)} & \colhead{(3)} & \colhead{(4)} & \colhead{(5)} & \colhead{(6)} & \colhead{(7)}
}
\startdata
1 & 18A-239 & 70 & 88 hrs & 18 & 11 & 7\\
2 & 19B-144 & 63 & 84 hrs & 4 & 1 & 3\\
3 & 20A-089 & 58 & 111 hrs & 8 & 2 & 6\\
4 & 20B-045 & 92 & 129 hrs & 31 & 19 & 12\\
5 & 21A-389 & 95 & 52 hrs & 47 & 40 & 9\\
\hline
Total & & 378 & 464 hrs & 110 & 73 & 37
\enddata
\tablecomments{col.(1): Campaign number.\ col.(2): GBT Program ID.\ col.(3): Total number of targets in the campaign.\ col.(4): Total observing time in the campaign.\ cols.(5-7): Number of \ion{H}{1} detections, LSB dwarfs, and UDGs in campaign.\ The final row lists the total for all campaigns.}
\end{deluxetable*}

\section{Observations and Data Reduction} \label{sec:Obsanddata}
We performed a total of $\sim464$ hours of position-switched \ion{H}{1} observations between 2018 February and 2021 August using the GBT along the lines of sight (LOS) to the 378 SMUDGes candidates across five observing campaigns.\ The integration time for each campaign along with other details are summarized in Table \ref{tab:surveysummary}.\

Our observational setup for Campaigns 2 and 3 follows that of our pilot survey/Campaign 1 described in \citetalias{2020Karunakaran}.\ Briefly, we used the L-band receiver coupled with the Versatile GBT Astronomical Spectrometer (VEGAS) set to Mode 7 which provides a spectral resolution of 3.1 kHz and bandpass of 100 MHz.\ This setup allows for the detection of \ion{H}{1} emission lines out to $V_{Helio} \sim 14000 \, \kms$.\ Due to the increased incidences of radio frequency interference (RFI) and to mitigate loss of data, we employed an alternative VEGAS setup for Campaigns 4 and 5.\ To mimic the velocity coverage of Campaigns 1-3, we take advantage of the sub-band modes of VEGAS which allow for simultaneous observations across multiple spectral ranges.\ Specifically, we use Mode 21 which provides a spectral resolution of 2.9 kHz and a bandpass of 23.44 MHz for each sub-band.\ We use three sub-bands to provide continuous coverage between $-1000 \, \kms \lesssim V_{Helio} \lesssim 13500 \, \kms$.\ 

As in Campaign 1, we estimate the integration times for all of our survey targets using $m_g$ to reach a gas richness of $\frac{\mhi}{\LG} \sim$ 1 $\frac{\msun}{\lsun}$ with $S/N = 5$ in a single $50 \, \kms$ channel.\ Gas richness is a distance-independent quantity in the nearby Universe since both $\mhi$ and $\LG$ scale with distance squared.\ As such, a single observation allows us to search for an \ion{H}{1} reservoir in our targets anywhere within our bandpass coverage.\ 

All of the raw data were reduced using the standard GBTIDL\footnote{http://gbtidl.nrao.edu/} procedure $getps$.\ We remove narrow-band and broadband RFI using the same procedures from \citetalias{2020Karunakaran}.\ We note that the primary source of broadband RFI for our observations is the intermittent GPS-L3 signal at $\sim1.381$ GHz.\ The overwhelming strength of this signal significantly affects the entire bandpass of our spectra.\ While removal of this effect is trivial via flagging individual 5s integrations in Campaigns 1-3, the alternative observational setup in Campaigns 4 and 5 would allow us to only flag integrations from the latter two sub-bands and leave us with usable data in the first sub-band.\ Once the RFI removal is complete, we smooth our spectra to our desired resolutions as in \citetalias{2020Karunakaran}.\ We note that we scale the fluxes in our final spectra up by 20\% to account for the systematic offset in the GBT noise diode calibration values reported by \citet[][]{gbtcal}.\ The RMS noise, $\sigma_{50}$, for each spectrum at $\Delta V = 50\,\kms$ resolution is given in column 13 of Table \ref{table:maintable}.\ Specifically for targets from Campaigns 4 and 5, we provide $\sigma_{50}$ using the first sub-band when a significant portion of their spectra have been flagged due to RFI and for all other targets list the median $\sigma_{50}$ value across the three sub-bands. 

We follow the same steps as in \citetalias{2020Karunakaran} to search for sources of statistically significant emission in our spectra.\ We examine the calibrated, RFI-excised spectra by-eye after smoothing to multiple resolutions from $5-50 \, \kms$.\ We detect \ion{H}{1} emission along the LOS to 110 SMUDGes candidates (column 14 of Table \ref{table:maintable}).\ We find no significant \ion{H}{1} emission associated with the 268 remaining targets and calculate stringent $5\sigma$ the upper limits on \ion{H}{1} mass, ${\mhilim}$, and gas richness, ${\mhilim}/{\LG}$.

We highlight that the detection rates $(\mathrm{N_{det}/N_{targets}})$ significantly increase with the inclusion of the \textit{GALEX} UV criterion from $30/191\sim15\%$ across Campaigns 1-3 to $31/92\sim33\%$ in Campaign 4.\ This rate increased further for Campaign 5 with the relaxed central surface brightness limit to $47/95\sim49\%$.\ These increases highlight the gain in detection efficiency using a star-forming criterion, whether through \textit{GALEX} UV or optical colors, as seen in previous \ion{H}{1} studies \citep[e.g.][]{GASS,Brown2015}.

\section{Results} \label{sec:results}
\subsection{Properties of \ion{H}{1} Detections} \label{subsec:detections}
We detect \ion{H}{1} along the LOS to 110 SMUDGes candidates.\ The GBT beam (FWHM $\sim \mathrm{9.1'}$ at $\sim$ 1.4 GHz) response is well understood down to $\approx -30 \mathrm{dB}$ with its first sidelobe occurring at $\sim0.35$ deg from the center of the beam \citep[e.g.,][]{GBTbeam}.\ To ensure that our \ion{H}{1} detections are associated with our SMUDGes candidates and not a nearby gas-rich interloper we search through NED\footnote{The NASA/IPAC Extragalactic Database (NED) is operated by the Jet Propulsion Laboratory, California Institute of Technology, under contract with the National Aeronautics and Space Administration.} and the DESI Legacy Imaging Survey Sky Viewer\footnote{http://legacysurvey.org/viewer} for objects within $30'$ of the LOS.\ Our search found no such interlopers for any of our \ion{H}{1} detections, and we conclude that they are the \ion{H}{1} counterparts to our SMUDGes candidates.\ 

We note, however, that during our search for interlopers, we found that a subset (36) of our detections have previously been detected in \ion{H}{1} studies.\ 34 SMUDGes candidates were detected in the ALFALFA $\alpha.100$ catalog \citep{Haynes2018}.\ Half of these objects (17/34) were not included in the \ion{H}{1}-bearing ultra-diffuse sources (HUDs) samples \citep{2017Leisman,2019Janowiecki} and we, therefore, present them here for the first time.\ Additionally, one of these candidates, SMDG0050362+322104, is only marginally detected in our GBT data within a relatively noisy spectrum and was more clearly detected in ALFALFA.\ We, therefore, use the ALFALFA-derived properties for this one object.\ One UDG candidate (SMDG0959362+114628) was detected as part of an \ion{H}{1} follow-up study of the MATLAS dwarf and UDG samples \citep[MATLAS-0607,][]{2022Poulain}.\ The final previously published UDG candidate (SMDG1306204+040902) was presented as part of the RESOLVE survey \cite[RS0955,][]{2023Hutchens}.\ We compare the derived \ion{H}{1} properties we derive below to those from these samples in Appendix \ref{sec:a100comp}, finding general consistency between them.\

To derive distance-independent quantities from the spectra (systemic velocity, $\vsys$, velocity width, $\wfty$, and \ion{H}{1} flux, $S_{HI}={\int}S{\delta V}$), we take two approaches.\ The first approach is described in \citetalias{2020Karunakaran} and follows the general procedure of \citet{2005Springob}.\ Briefly, we fit a first-order polynomial to each edge of the \ion{H}{1} profile between 15\% and 85\% of the peak flux value.\ We then find the velocities corresponding to the 50\% flux value for each edge and take their mean as $\vsys$ and their difference as $\wfty$.\ The \ion{H}{1} flux, $S_{HI}={\int}S{\delta V}$, is determined by taking the integral of the \ion{H}{1} signal.\ Our second approach involves modeling the spectra with the Busy function \citep{2014Westmeier} using the \texttt{BusyFit} software\footnote{\url{https://www.atnf.csiro.au/people/Tobias.Westmeier/tools_software_busyfit.php}} which also outputs the aforementioned distance-independent quantities.\ We take the mean of the quantities calculated from the two approaches as our fiducial values and their differences as the respective uncertainties.\ We note that the differences in these values are marginally larger than if we were to propagate the uncertainties on these quantities from each approach in quadrature and use the former as a more conservative error estimate.\ We further correct $\wfty$ for instrumental and cosmological redshift broadening following \citet{2005Springob}, $\wftyc$, as well as ISM turbulence following the methods of \citet{2001Verheijen}, $\wftyct$, as in our previous work \citepalias[i.e.][]{2020Karunakaran} and list $\wftyct$ in column 5 of Tables \ref{table:detectiontable} and \ref{table:dwarftable}.\ 

Arguably the most important quantity for our SMUDGes candidates is an estimate of their distance.\ We use our derived $\vsys$ and the Hubble-Lema\^{i}tre Law to estimate kinematic distances for all of our \ion{H}{1} detections and adopt a distance uncertainty of 5 Mpc \citep{2017Leisman, HCGUDGs}.\ Using these distances and the angular sizes, $r_{\mathrm{eff}}$, from Table \ref{table:maintable}, we confirm 37 UDGs with $R_{\mathrm{eff}} > 1.5\,\mathrm{kpc}$ and $\mu_{0,g}\gtrsim 24 \,\mathrm{mag\,arcsec^{-2}}$, and give their \ion{H}{1} properties in Table \ref{table:detectiontable}.\ The remaining detections are, therefore, low-surface brightness (LSB) dwarf galaxies and their \ion{H}{1} properties are listed in Table \ref{table:dwarftable}.\ This sample, to the best of our knowledge, marks the largest targetted \ion{H}{1} follow-up sample of optically-detected UDGs to date.\ It also provides a unique context for the UDG population as it contains galaxies that are both equally large but marginally higher surface brightness and equally low surface brightness but physically smaller.

With distance estimates in hand, we now estimate two crucial physical properties: \ion{H}{1} masses, $\mhi$, and stellar masses, $M_{*}$.\ To calculate $\mhi$, we use our \ion{H}{1} fluxes and our kinematic distances to determine using the standard equation for an optically thin gas \citep{1984Haynes}: \begin{equation} \label{eqn:himass} M_{HI}=2.356\times 10^{5}(D_{HI})^{2}S_{HI} \, \mathrm{\msun},\end{equation} where the distance, $D_{HI}$, is in Mpc and $S_{HI}$ is in Jy $\kms$ and their uncertainties are calculated following \cite{2005Springob}.\ We calculate $M_{*}$ for our detections using $m_{g}$ and $g-r$ \citepalias[with bias corrections applied, see][for details]{2021Zaritsky,2022Zaritsky} from Table \ref{table:maintable} in the relations of \citet[][]{2017zhang} and assuming $D_{HI}$.\ Since photometric properties for our targets have asymmetric uncertainties \citepalias[see][]{2021Zaritsky,2022Zaritsky}, we do not propagate uncertainties on any of these quantities (e.g.\ $g-r$ and $M_{*}$) in quadrature and, instead, make use of the \texttt{asymmetric$\_$uncertainty} Python package \citep{2022gobat} which implements the methodology presented in \citet{2008Starling}.\ This allows us to estimate asymmetric uncertainties on any additional quantities in this work, notably the ratio of \ion{H}{1} and stellar mass or gas-richness, $\mhi/M_{*}$, the physical effective radius, $R_{\mathrm{eff}}$, and the baryonic mass, $\mbary=1.33\mhi+M_{*}$.\ All of these distance-dependent quantities are listed for our UDGs and LSB dwarfs in Tables \ref{table:detectiontable} and \ref{table:dwarftable}, respectively.\

\begin{figure*}[htb!]
\includegraphics[width=18cm]{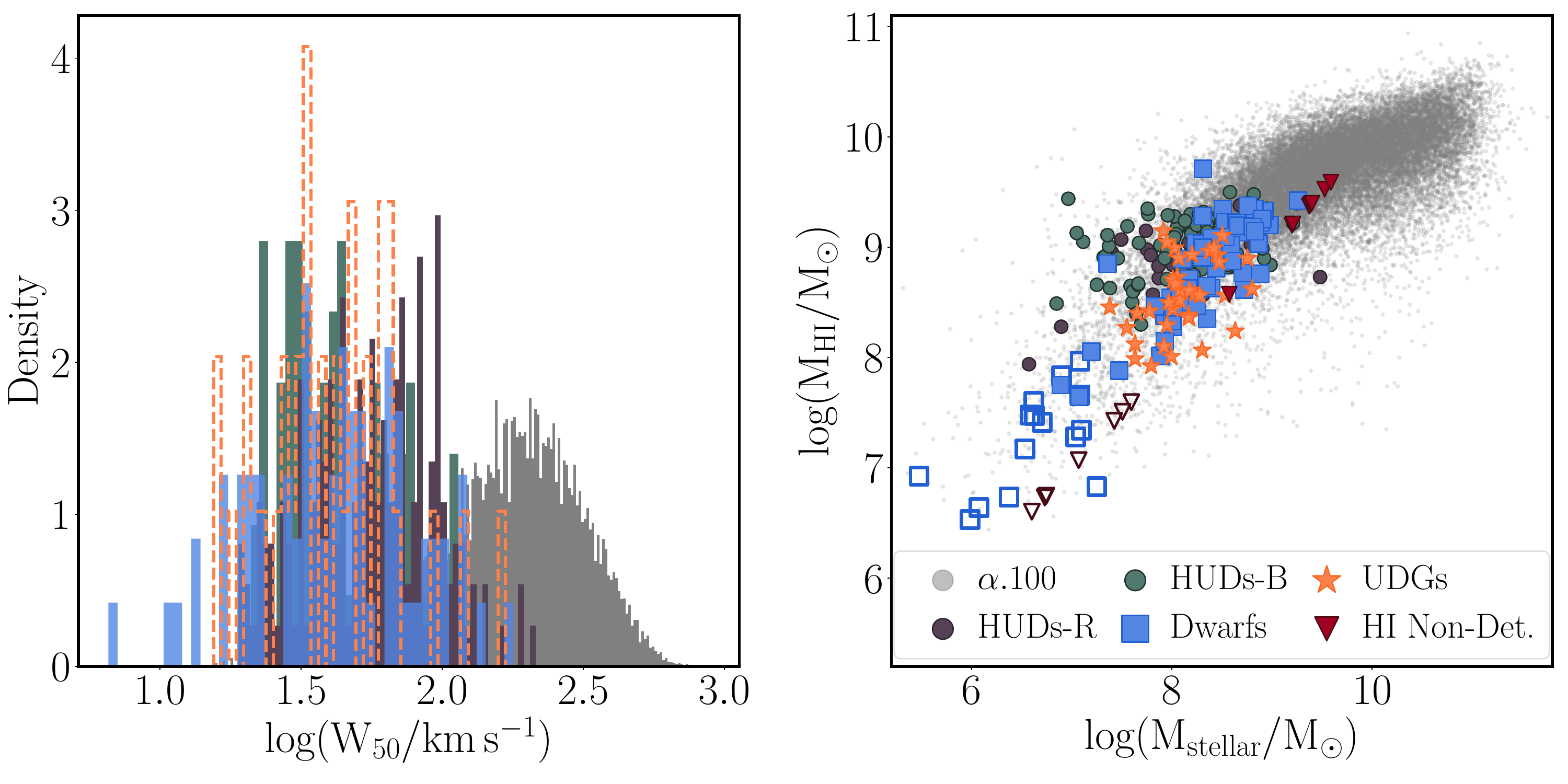}
\caption{Comparison of derived properties between our \ion{H}{1} detections, the \ion{H}{1}-bearing Ultra Diffuse Sources (HUDs), and the ALFALFA sample.\ $Left:$ Distribution of velocity widths $\wftyct$ for SMUDGes UDGs (orange), SMUDGes LSB dwarfs (blue), the HUDs-B and -R samples (purple and green, respectively), and galaxies from the $\alpha.100$ catalog with SDSS and GALEX coverage (grey).\ $Right:$ $\mhi - M_{*}$ relation for the same samples as in the left panel.\ We also show our \ion{H}{1} non-detections with distances estimates as inverted triangles.\ The open symbols in the right panel indicate sources that would not be included in HUDs due to their low recessional velocities.}
\label{fig:hipropcomp}
\end{figure*}

In Figure \ref{fig:hipropcomp}, we compare the derived properties of our \ion{H}{1}-confirmed UDGs (orange) and LSB dwarfs (blue) to those from the HUDs sample (HUDs-B in green and HUDs-R in purple) and the ALFALFA $\alpha.100$ catalog \citep[shown in gray,][]{2018Haynes}.\ We compare the distributions of $\wftyct$ in the left panel and, as seen in previous studies of UDGs (and LSB dwarfs) in \ion{H}{1} \citep[e.g.][]{2019Janowiecki,2022Poulain}, the \ion{H}{1} detections from our survey predominantly have narrow velocity widths and are generally more similar to the HUDs sample than the broader ALFALFA sample.\ In the right panel of Figure \ref{fig:hipropcomp}, we show the $\mhi-M_{*}$ relation.\ Both the UDGs (orange stars) and LSB dwarfs (blue squares) in our sample span a similar space in this plane as the HUDs (green and purple circles) and fall towards the low-mass end of the ALFALFA sample \citep[gray points][]{2020Durbala}.\ The empty square symbols are a subset of our sources with distances that fall below the 25 Mpc limit set for the HUDs sample due to potentially large distance uncertainties.\ As noted in \citetalias{2020Karunakaran}, our sample is marginally less gas-rich when compared to the HUDs sample, potentially highlighting the differences in targeted and untargeted (i.e. ``blind'') surveys.\

\begin{figure*}[htb!]
\includegraphics[width=18cm]{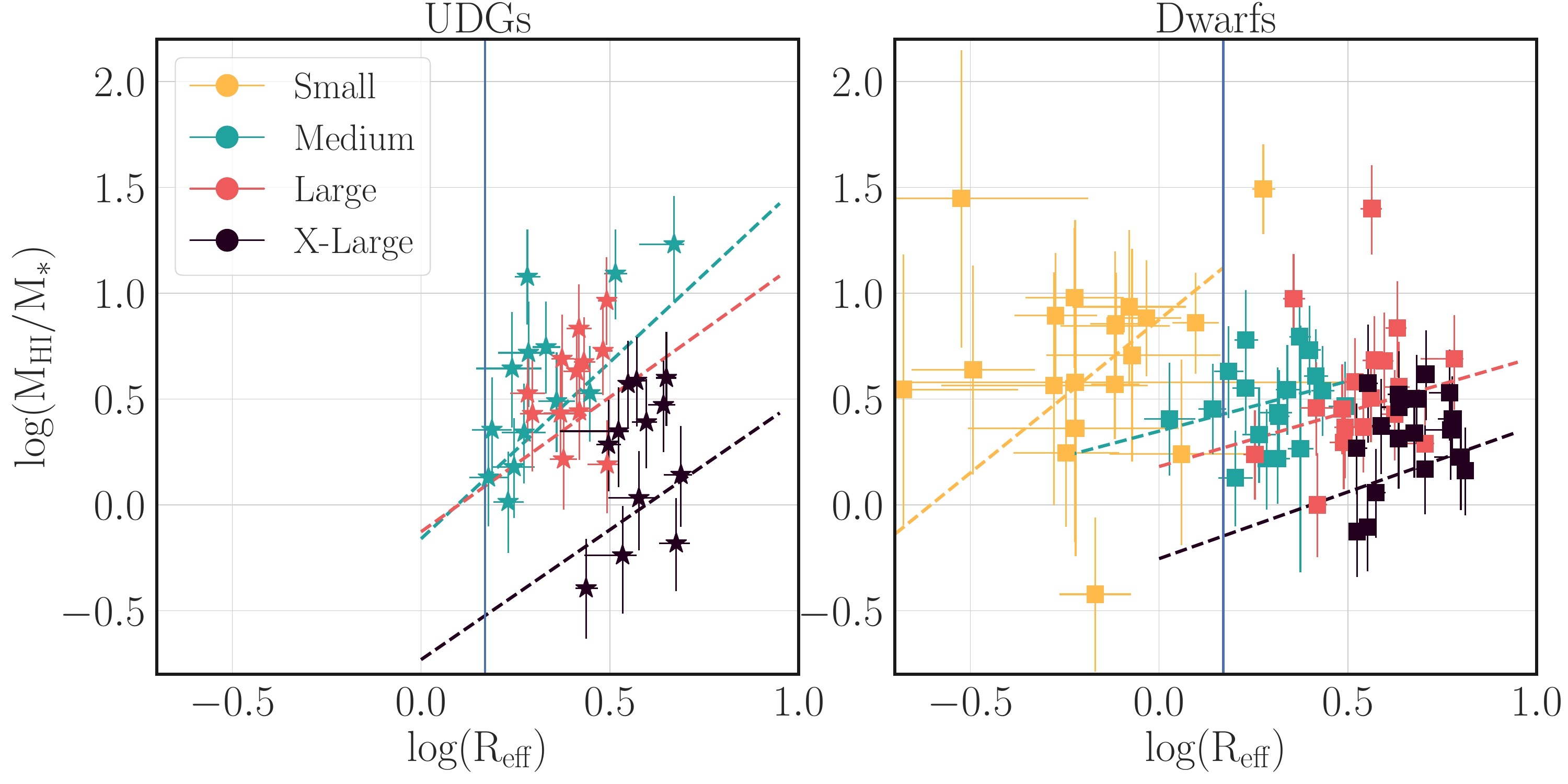}
\caption{Gas-richness, $\mathrm{log(\mhi/\mstar)}$, as a function of physical size, $\mathrm{log(R_{\mathrm{eff}})}$, for \ion{H}{1}-confirmed UDGs (left) and LSB dwarfs (right).\ The symbol colors correspond to their stellar mass bins, defined to populate the bins with roughly equal numbers of objects.\ LSB dwarfs: Small (yellow) - $\mathrm{log(\mstar)}= [5.47,7.35)$, Medium (green) - $\mathrm{log(\mstar)}= [7.35,8.30)$, Large (pink) - $\mathrm{log(\mstar)}= [8.30,8.59)$, X-Large (black) - $\mathrm{log(\mstar)}\geq8.59$. UDGs: Medium (green) - $\mathrm{log(\mstar)}= [7.37,7.99)$, Large (pink) - $\mathrm{log(\mstar)}= [7.99,8.20)$, X-Large (black) - $\mathrm{log(\mstar)}\geq8.20$.\ The vertical line in both panels corresponds to the ${R_{\mathrm{eff}}}>1.5$ kpc criterion.\ The dashed colored lines in both panels correspond to the best-fit relations in this plane for each subset obtained using \texttt{LinMix}.}
\label{fig:gasrichness-size}
\end{figure*}

We show the relationship between gas-richness, $\mhi/M_{*}$, and physical optical size, $R_{\mathrm{eff}}$, for our \ion{H}{1}-confirmed UDGs (left panel) and LSB dwarfs (right panel) in Figure \ref{fig:gasrichness-size}.\ We have divided the galaxies in each panel into stellar mass bins with approximately equal numbers.\ The vertical blue line in each of the panels represents the $R_{\mathrm{eff}}=1.5\,\mathrm{kpc}$ size criterion for UDGs.\ For each subset, we obtain a best-fit linear relation between $\mathrm{log}[\mhi/M_{*}]$, and $\mathrm{log}[R_{\mathrm{eff}}]$ using the Python implementation of \texttt{LinMix}\footnote{\url{https://github.com/jmeyers314/linmix}} \citep{2007Kelly}, a hierarchical Bayesian fitting routine.\ We note that we have assumed symmetric uncertainties for this procedure, selecting whichever is larger of the lower and upper bounds on these derived quantities.\ The best-fit relations are shown for all stellar mass bins for UDGs and LSB dwarfs as dashed lines with colors matching their respective bins.\ 

Overall, Figure \ref{fig:gasrichness-size} shows that within a given stellar mass bin, physically larger galaxies tend to have larger gas richness.\ Between the UDGs and LSB dwarfs, the former have steeper relations between $\mathrm{log}[\mhi/M_{*}]$, and $\mathrm{log}[R_{\mathrm{eff}}]$.\ However, the scatter in the best-fit parameters for these relations is large and we show them in more detail in the appendix in Figure \ref{fig:fitcomp_approaches}.\ We only find a significant difference in the Medium stellar mass bin between the dwarf and UDG samples, with the UDGs having a significantly steeper slope ($\mathrm{median\pm interquartile\,range, IQR} = 2.3^{+0.6}_{-0.7}$) compared to the LSB dwarfs ($\mathrm{median\pm IQR} = 0.4^{+0.5}_{-0.4}$).\ 

To explore the robustness of these fitting results, we take two similar, yet alternative, approaches to characterizing the relationships in this plane and describe them in more detail in Appendix \ref{sec:AltFit}.\ Briefly, we perform bootstrapped orthogonal distance regression and Monte Carlo sampling from a Split-Normal distribution.\ We find no significant differences in the results from these two methods and find similarly large scatter as before.\ Similarly, we find that restricting our investigation to only large (i.e.\ $R_{\mathrm{eff}}>1.5$ kpc) dwarf galaxies only marginally shifts our results and, again, only the Medium stellar mass bin has a statistically significant difference between dwarf and UDGs.\ Overall, we conclude that there is no significant difference between the majority of the UDGs and LSB dwarfs from our sample in the $\mathrm{log}[\mhi/M_{*}]$-$\mathrm{log}[R_{\mathrm{eff}}]$ plane.\ We briefly return to this result in Section \ref{subsec:udgdef}.\

\subsection{\ion{H}{1} Non-detections} \label{subsec:HInondetect}
We find no statistically significant \ion{H}{1} emission along the LOS to the vast majority (268/378) of our targeted SMUDGes candidates.\ For these sources, we smooth their spectra to $\Delta V = 50\, \kms$ and measure a representative RMS noise, $\sigma_{50}$.\ In our pilot study focusing on the Coma Cluster region, we could readily place stringent, $5\sigma$ \ion{H}{1}-mass upper limits by assuming the cluster's distance.\ However, given the broad extent of the full survey sample, we instead place (distant-independent) upper limits on the ratio of \ion{H}{1}-mass to $g-$band luminosity, $\mhilim/\LG$.\ Both $\sigma_{50}$ and $\mhilim/\LG$ are listed in Table \ref{table:maintable}.\

A subset of our \ion{H}{1} non-detections have distance and/or redshift measurements available in the literature.\ We make note of these in Table \ref{table:maintable} and briefly discuss them here.\ As noted in \citetalias{2020Karunakaran}, SMDG1221577+281436 was reported as the marginal \ion{H}{1} detection by \citet[][d1221+2814]{huch2009} but our deeper GBT data found no such emission.\ Similarly, in \citetalias{2020Karunakaran} we noted that SMDG1302417+215954 (IC 4107) is a known \ion{H}{1} non-detection \citep{1992schombert} and we adopted a distance estimate ($D_{lim}=3.8\,\mathrm{Mpc}$) based on its SDSS spectrum \citep[$V_{opt} = 267\,\kms$][]{2014KimExVirgo} and its likely membership within NGC 4826 group \citep{2014Karachentsev}.\ Several targets have optical velocities from \citet{2021Kadowaki} as measured with data from the Large Binocular Telescope: SMDG0244338-001601 ($1282\,\kms$), SMDG1217378+283519 ($790\,\kms$), SMDG1221086+292920 ($1316\,\kms$), SMDG1237294+204442 ($1751 \, \kms$), SMDG1240530+321656 ($7077\,\kms$), SMDG1242314+315809 ($826 \, \kms$), SMDG1245276+181803 ($1823 \, \kms$), SMDG1251013+274753 ($6406 \, \kms$), SMDG1253151+274115 ($7592\,\kms$), and SMDG1257017+282325 ($6864 \, \kms$).\ \citet{2018RuizLara} measure an optical velocity for SMDG1300206+274712 of $V_{sys} = 6548 \, \kms$ using data from OSIRIS on the Gran Canarias Telescope.\ SMDG1300580+265835 has two measured velocities from \citet{2015vandokkumb}, $V_{sys} = 6280 \, \kms$, using data from LRIS on the Keck Telescope, and from \citet{2018Gu}, $V_{sys} = 6402 \,\kms$, using data from the SDSS-IV MaNGA survey.\ \citet{2018Gu} also measure a velocity of $V_{sys} = 8315 \,\kms$ for SMDG1301582+275011.\ SMDG1429098+331038 has a measured velocity of $V_{sys} = 898 \,\kms$ from the AGN and Galaxy Evolution Survey \citep{2012Kochanek}.\ As part of the Exploration of Local Volume Satellites (ELVES) survey, \citet{2022Carlsten} estimate surface brightness fluctuation (SBF) distances to their dwarf satellite galaxy candidates.\ They estimate an SBF distance of $D_{sbf}=9.48$ Mpc for SMDG1119215+140432 (dw119+1404) and also list a recessional velocity of $V_{sys}=486\,\kms$.\ Finally, the recessional velocity of SMDG0220482-002754 was determined to be $V_{sys} = 1395 \,\kms$ using Keck Cosmic Web Imager observations as part of a follow-up program of detections from the Dragonfly Ultrawide Survey \citep[][]{2024Shen}.\ We summarize all of these references in Table \ref{table:maintable}.\

\begin{figure*}[htb!]
\includegraphics[width=18.5cm]{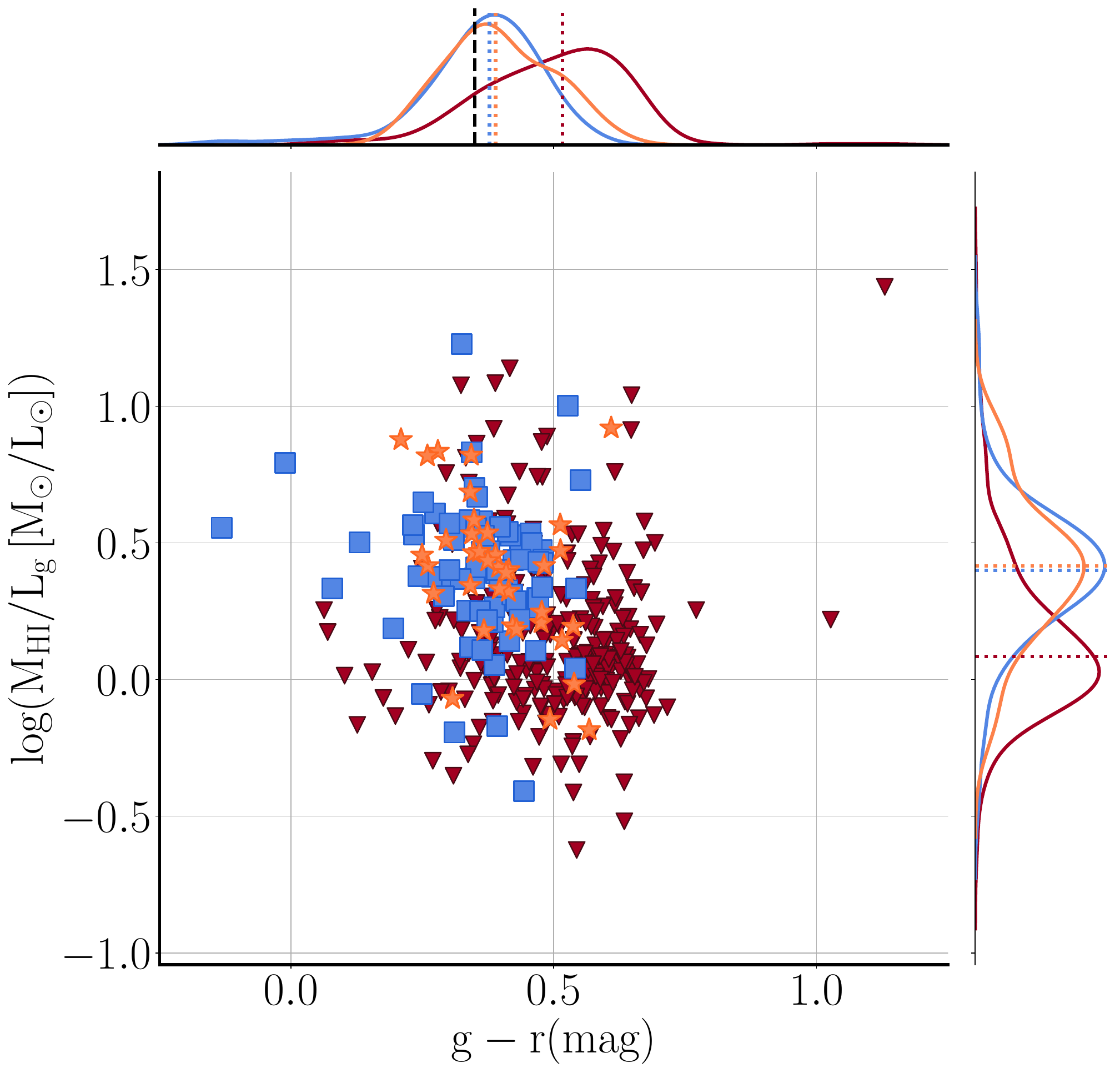}
\caption{$\mhi/\LG$ (orange stars and blue squares for \ion{H}{1}-confirmed UDGs and LSB dwarfs, respectively) and $\mhilim/\LG$ (red inverted triangles for non-detections) as a function of $g-r$ color for our sample.\ The marginalized distributions are shown along the opposite axes and we show the median of each subset as dotted lines.\ For comparison in the $g-r$ distribution, we show the median $g-r=0.35$ color of the HUDs sample \citep{2017Leisman} as the dashed black line.}
\label{fig:gasrichness-colour}
\end{figure*}

\begin{figure*}[htb!]
\includegraphics[width=18cm]{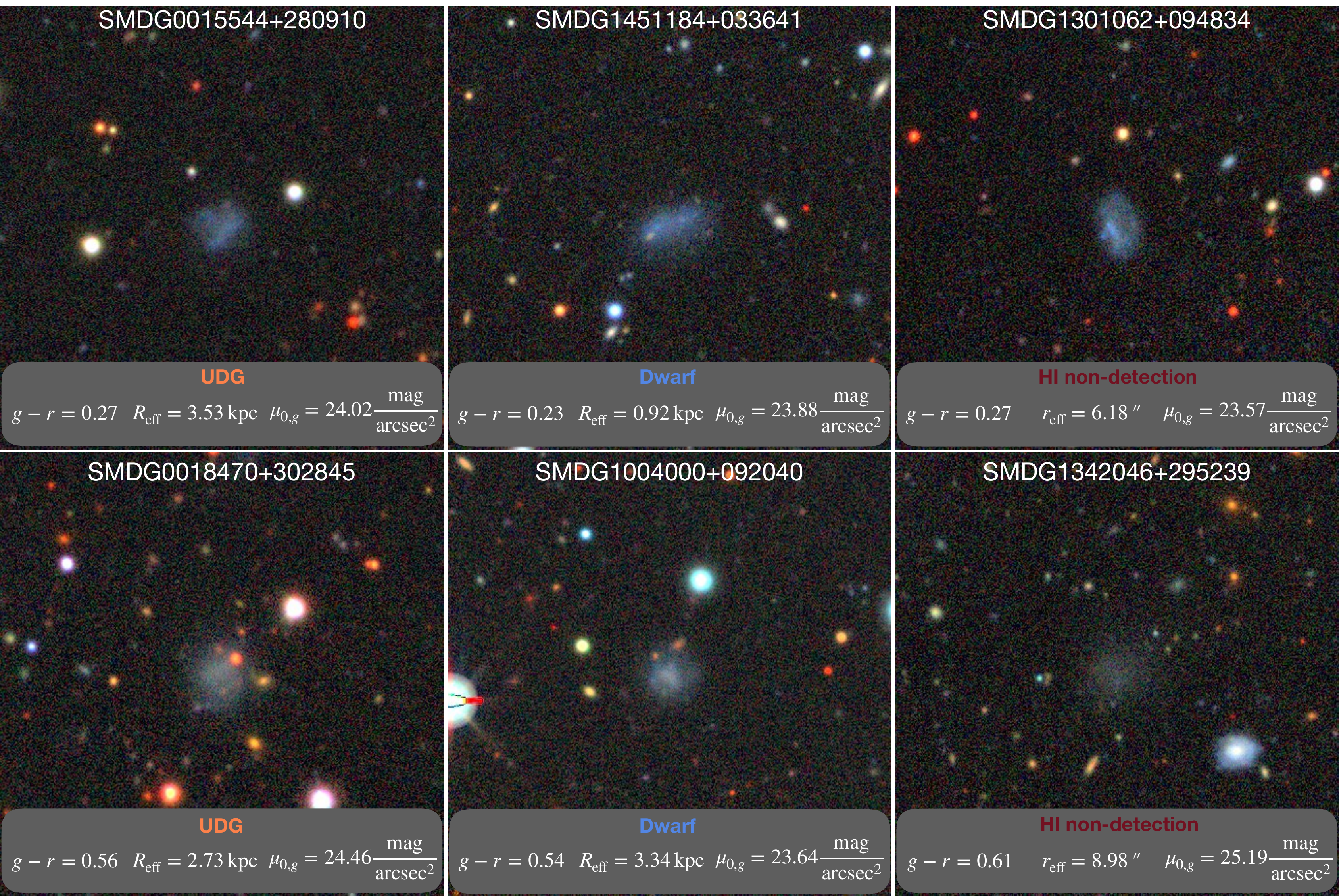}
\caption{$\sim2\mathrm{arcmin}\times2\mathrm{arcmin}$ color $grz$ image cutouts of example galaxies from our sample.\ We show two randomly selected examples of \ion{H}{1}-confirmed UDGs (left column), \ion{H}{1}-confirmed LSB dwarfs (middle column), and \ion{H}{1} non-detections (right column).\ In each pair, one was chosen to fall towards the blue end of the $g-r$ color distribution (top row) and the other towards the red end (bottom row).\ For the UDGs and LSB dwarfs, we list their colors ($g-r$ ), physical effective radii ($R_{\mathrm{eff}}$), and central surface brightness ($\mu_{0,g}$).\ For our non-detections, we show the same values but instead list their angular effective radii ($r_{\mathrm{eff}}$).}
\label{fig:cutouts}
\end{figure*}

We briefly compare the \ion{H}{1} properties of our \ion{H}{1} detections and non-detections.\ For the subset of non-detections with distance estimates $(D_{lim})$, we estimate 5$\sigma$ \ion{H}{1} mass upper limits using a modified version of Eq.\ \ref{eqn:himass}: \begin{equation} \label{eqn:himasslim} M_{HI,lim}=2.356\times 10^{5}(D_{lim})^{2}(5\sigma_{50})(\Delta V) \, \mathrm{\msun},\end{equation} assuming a velocity width of $\Delta V=50\kms$ and $\sigma_{50}$ from Table \ref{table:maintable}.\ We plot these limits alongside our \ion{H}{1} detections in Figure \ref{fig:hipropcomp}.\ It is evident that these \ion{H}{1} limits are typically lower at a given stellar mass compared to our \ion{H}{1} detections.\

We show $\mhi/\LG$ for our \ion{H}{1} detections and $\mhilim/\LG$ for our \ion{H}{1} non-detections (red inverted triangles) as a function of their $g-r$ colors.\ Along the top and right sides of the figure, we show the marginalized distributions of these quantities for each of the subsets.\ We also highlight the median within each of these samples with correspondingly colored dashed lines.\ We find a similar result, with respect to the \ion{H}{1} mass upper limits, as in \citetalias{2020Karunakaran}, where measured values are higher than those used to estimate required integration times.\ In fact, there are a handful of more extreme cases where $\mathrm{log}(\mhilim/\LG) >0.5$.\ Overall, we find similar reasoning for this apparent decrease in sensitivity (i.e.\ integrations/scans flagged due to RFI and 20\% calibration adjustment).\ However, in our investigation of these cases, we found their spectra to be significantly noisier than expected.\ We nevertheless opt to provide estimates of $\mhilim/\LG$.\ 

The \ion{H}{1} detections in our sample tend to have lower (i.e.\ bluer) $g-r$ colors (see Figure \ref{fig:cutouts} for a visual comparison) and are more gas-rich (i.e.\ larger $\mhi/\LG$).\ This can be seen more clearly in the marginalized distributions along with median values: $(g-r)_{\mathrm{UDG}} = 0.39$, $(g-r)_{\mathrm{Dwarf}} = 0.38$, $(g-r)_{\mathrm{Nondet}} = 0.52$, $\mathrm{log}(\mhi/\LG)_{\mathrm{UDG}} = 0.42$, $\mathrm{log}(\mhi/\LG)_{\mathrm{Dwarf}} = 0.40$, and $\mathrm{log}(\mhilim/\LG)_{\mathrm{Nondet}} = 0.52$.\ Furthermore, for comparison, we also show the median $g-r=0.35$ of the entire HUDs sample as the vertical dashed-dotted line.\ In Figure \ref{fig:cutouts}, we show visual examples of \ion{H}{1} detections and non-detections using image cutouts from the Legacy Survey.\ The top and bottom rows show optically blue and red systems, respectively, while the left, middle, and right columns show UDGs, LSB dwarfs, and \ion{H}{1} non-detections, respectively.\ While our \ion{H}{1} detections are typically bluer, there are clearly some exceptions.\ We compare the global properties between our \ion{H}{1} detections and non-detections further in Section \ref{subsec:gasrich}.\

\subsection{Star formation and Environment}\label{sec:sfrenv}
In addition to the \ion{H}{1} properties we presented above, we have also characterized the star formation and environments of our \ion{H}{1} detections.\ These results and our methodologies will be described in more detail in a subsequent paper (Karunakaran et al., in prep) and we briefly describe them here.\

We perform aperture photometry for our entire sample using archival \textit{GALEX} UV imaging.\ We searched through MAST for the deepest available imaging in both NUV and FUV.\ Across our sample, 339 have both NUV and FUV data and 18 only have NUV data.\ Here, we focus only on the properties of our \ion{H}{1} detections and briefly discuss those for our \ion{H}{1} non-detections in Section \ref{subsec:gasrich}.\ To estimate the UV fluxes, we follow a similar methodology to \citet{2022Carlsten} and perform aperture photometry using an aperture that is twice the size of the optical effective radius, $2r_{\mathrm{eff}}$, and opt to estimate the local background and noise using an annulus between $2r_{\mathrm{eff}}$ and $3r_{\mathrm{eff}}$.\ We compared the results from this approach to the curve-of-growth and random sampling method we applied in previous work \citep[i.e.][]{2021Karunakaran} with a small subset from our sample and found consistent results.\ We considered a source to have a detection if the total S/N $>2$.\ 

In total, 88 of our \ion{H}{1} detections have both NUV and FUV detections, 12 have only NUV detections, three are not detected in either band, and seven have no available data.\ Using these fluxes and the standard relations from \citet{2007Morrissey}, we calculate their apparent magnitudes and correct them for foreground extinction.\ We then assume the distances derived from their \ion{H}{1} emission to calculate their UV luminosities and estimate star formation rates (SFRs) using the relations of \citet{2006IglesiasParamo}.\

\begin{figure*}[htb!]
\includegraphics[width=18.5cm]{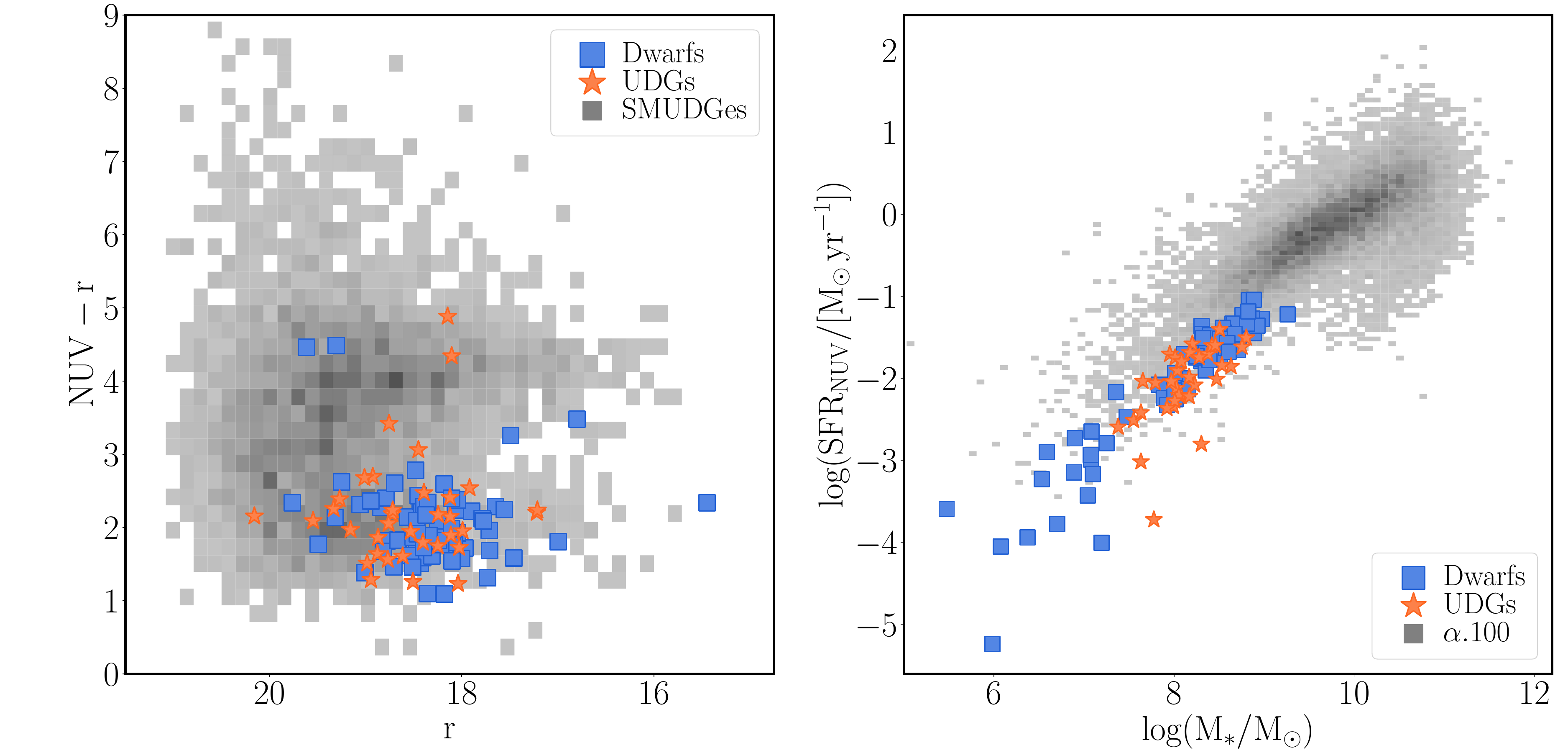}
\caption{($\mathrm{NUV}-r$) vs.\ r (left) and NUV star formation rate (SFR) as a function of stellar mass (right) for our sample.\ Once again, the orange stars and blue squares for \ion{H}{1}-confirmed UDGs and LSB dwarfs, respectively.\ We show the NUV-IR SFRs from \citet{2020Durbala} for the $\alpha.100$ catalog.\ The SFRs from our sample tend to fall towards the lower end of the $\alpha.100$ distribution and appear to have specific SFRs ($\mathrm{SFR}/\mstar$) below the main sequence from the $\alpha.100$ population.\ }
\label{fig:sfr_mstar}
\end{figure*}

In Figure \ref{fig:sfr_mstar}, we show ($\mathrm{NUV}-r$) vs.\ r in the left panel and the SFR-$\mstar$ plane in the right panel.\ Our UDGs and LSB dwarfs are shown as orange stars and blue squares respectively.\ In the left panel, we also include the NUV photometry for the complete SMUDGes sample from Zaritsky et al. (in prep) as the gray histogram\footnote{We note that the different methods for UV photometry used in this paper and for the complete SMUDGes sample are consistent with one another, particular for those with \ion{H}{1} detections.}.\ As expected, the majority of our \ion{H}{1} detections fall toward the lower end of the $\mathrm{NUV}-r$ distribution and overlap with the ``blue'' cloud of the SMUDGes sample.\
In the right panel, for comparison, we instead plot the SFRs for the $\alpha.100$ sample which were derived using \textit{GALEX} NUV imaging corrected for infrared (IR) emission using \textit{WISE}, \citep[see][for more details]{2020Durbala}.\ The majority of our sample tend to have lower estimated SFRs at a given stellar mass compared to the galaxies from $\alpha.100$ and are lower than those estimated for the HUDs sample \citep[see Figure 6 in][]{2017Leisman}.\ We note that the IR corrections at low stellar masses should not result in significant changes in their estimated SFRs unless they are undergoing a starburst \citep[i.e.][]{2015McQuinn}.\ The few outliers from our \ion{H}{1} sample that have $\mathrm{NUV}-r\gtrsim3$ and $m_{r}\gtrsim18$ correspond to the same outliers that have relatively low SFRs in the SFR-$\mstar$ plane.\ We discuss the implications of these star-forming properties more in Section \ref{subsec:formation}.\

To assess the broader environment of our \ion{H}{1} detections, we search through NED for any galaxies that have a $|\Delta\vsys| < 500\,\kms$ within a projected separation of 1.5 Mpc.\ Through this search, we determined that 36 (12 UDGs and 24 LSB dwarfs) of our \ion{H}{1} detections have at least one nearby neighbor while the remaining 74 (25 UDGs and 49 LSB dwarfs) reside in relatively isolated environments.\ Focusing more closely on their local environment, following the criteria from \citet{2021Kadowaki} for nearby galaxies that have $|\Delta\vsys| < 300\,\kms$ within a projected separation of 500 kpc, we find that 15 of our sources (four UDGs and 11 LSB dwarfs) have at least one ``local'' neighbor.\ For our \ion{H}{1} non-detections with distance estimates, this search finds 14/16 have at least one nearby neighbour and 10/16 have at least one ``local'' neighbour.

Performing a similar search for our remaining non-detections is less straightforward given the lack of distance information for the vast majority.\ Nevertheless, we perform the same search for local neighbors and assume that they reside at the same distance (except for those with known distances, see above) finding 199 of our non-detections have at least one neighbor within a projected separation of 500 kpc.\ The results from this relatively simple environmental analysis follow the expected trends seen within other galaxy samples: optically red/quenched, gas-poor galaxies tend to reside in more dense environments compared to blue/star-forming, gas-rich galaxies \citep[e.g.][]{2017browngasstripping,2019prole,2021Tanoglidis}.\

\section{Discussion} \label{sec:Discussion}
In the following sections, we discuss similarities and differences between our \ion{H}{1} detections and non-detections (Section \ref{subsec:gasrich}) as well as LSB dwarfs and UDGs (Section \ref{subsec:dwvudg}), explore the constraints on UDG formation mechanisms that arise from our \ion{H}{1} observations (Section \ref{subsec:formation}), and briefly address how UDG definitions affect our results (Section \ref{subsec:udgdef}).

\subsection{\ion{H}{1} Detections vs.\ Non-detections} \label{subsec:gasrich}
We compare four aspects between our \ion{H}{1} detections and non-detections: \ion{H}{1} properties, star-forming properties, environments, and optical properties.\ 

In Section \ref{subsec:HInondetect} and Figure \ref{fig:gasrichness-colour}, we showed that our \ion{H}{1} detections are typically more gas-rich (i.e.\ higher $\mhi/\LG$) and bluer (i.e.\ lower $g-r$) in color than our non-detections.\ However, there are a couple of notable subsets in this comparison.\ First, a non-negligible number of non-detections are bluer than the median $g-r$ color of our \ion{H}{1} detections (see top-right of Figure \ref{fig:cutouts}).\ Furthermore, there are several very blue (i.e.\ $g-r<0.2$) \ion{H}{1} non-detections that have relatively stringent upper limits, $\mhilim/\LG$.\ One interpretation for blue, gas-poor systems is that they recently experienced star formation episodes that have expelled their \ion{H}{1} reservoirs, rendering any remaining \ion{H}{1} content well below our detection limit.\ For a few of these systems, it is also possible that their \ion{H}{1} emission is obscured by that of a nearby gas-rich neighbor where we see broader \ion{H}{1} signals.\ The second notable subset is our relatively red \ion{H}{1} detections that fall toward near or beyond the median $g-r$ color of our non-detections (see bottom-left and middle of Figure \ref{fig:cutouts}).\ One possible interpretation for these systems is that they had a significant episode of star formation early in their evolution which expelled their gas, leading to an older (i.e.\ redder) stellar population, and eventually re-accreted the expelled \ion{H}{1} content.\ We revisit the two mentioned interpretations in the context of formation scenarios in the following subsection.

We move on to briefly discuss the star-forming properties of our sample.\ As mentioned in Section \ref{sec:sfrenv}, 88 of our \ion{H}{1} detections have FUV and NUV detections, 12 are detected in NUV only, three are undetected in either band, and the remaining seven do not have \textit{GALEX} imaging.\ On average, there appears to be no significant difference in the derived SFRs between our UDGs and LSB dwarfs as seen in Figure \ref{fig:sfr_mstar}.\ Turning towards our \ion{H}{1} non-detections, we find that 110 have detections in both \textit{GALEX} bands, 54 are detected in NUV only, 90 have no UV detections, and 14 have no available data.\ It is unsurprising to see a large number of detections, primarily due to the introduction of our additional UDG candidate selection criteria for Campaigns 4 and 5.\ Indeed, 98/164 of the \ion{H}{1} non-detections with detections in FUV and/or NUV are from the latter two campaigns.\ The majority (87/90) of the UV non-detections are from the first three campaigns.\ Interestingly, our \ion{H}{1} detections with any UV detections have a higher median $\mathrm{log}[\mhi/\LG] \sim 0.40$ compared to our \ion{H}{1} non-detections with any UV detections $\mathrm{log}[\mhilim/\LG] \lesssim 0.1$.\ While our data may not have been sensitive enough to detect the lower-than-expected gas reservoirs of these \ion{H}{1}-poor, star-forming SMUDGes candidates, there is also a possibility that the \ion{H}{1} emission for a small number of candidates could be masked by that of a nearby source.\

In terms of the spatial distribution of our sample, the majority of our \ion{H}{1} detections are in relative isolation while the majority of our non-detections appear to have at least one nearby neighbor in projection as we described in Section \ref{sec:sfrenv}.\ This result is not obvious based on the qualitative perspective from their sky distributions in Figure \ref{fig:skydist}.\ Our findings are generally consistent with previous studies which have investigated the environmental dependence of gas content for galaxies at higher \citep{2017browngasstripping} and lower surface brightness \citep{2019Janowiecki}.\

The final item we consider is the optical properties of our \ion{H}{1} detections and non-detections.\ In \citetalias{2020Karunakaran}, we demonstrated that there are distinct qualitative differences in the morphologies of our \ion{H}{1} detections vs.\ our non-detections.\ We perform a similar visual inspection using our full sample and reach a similar conclusion: \ion{H}{1} detections appear to have more irregular morphologies with a clumpier nature owing to distinct regions of star formation, while \ion{H}{1} non-detections tend to have smoother morphologies (see Figure \ref{fig:cutouts}).\ We note, however, that we do not include the SMUDGes candidates with \ion{H}{1} non-detections from Campaigns 4 and 5 in this qualitative comparison because they were specifically selected to have signs of recent star formation.\ 

\subsection{LSB Dwarfs vs.\ UDGs}\label{subsec:dwvudg}
We have established that the LSB dwarfs and UDGs in our sample have similar \ion{H}{1} and apparent properties throughout this paper.\ Additionally, their global properties, e.g.\ $\mstar \,\mathrm{and\,} \mhi$, fall within the scatter of the general \ion{H}{1}-detected population.\ Two differences are their lower projected velocity widths, $\wfty$, and SFRs relative to the broader population (see Figures \ref{fig:hipropcomp}-left and \ref{fig:sfr_mstar}).\ The fact that these differences are present for our UDGs and LSB dwarfs suggests that the deviations may result from selection effects instead of the galaxies themselves.\ 

These differences make statistical comparisons to scaling relations, such as the baryonic Tully-Fisher relation (BTFR), difficult without additional data.\ This is especially true for studying the BTFR, which has garnered attention for significant reported deviations for some UDGs \citep{2019HUDsBTFR,2020ManceraPina}.\ The low $\wfty$ and relatively high $\mbary$ for our LSB dwarfs and UDGs imply that, if we were to apply the methodology typically used for high surface brightness galaxies to derive rotation velocities from \ion{H}{1} widths and optical axial ratios \citep[e.g.][]{2016Bradford}, then both our UDG and LSB samples would appear to deviate from the BTFR.\ However, this approach is unreliable for stochastically star-forming dwarfs selected by surface brightness (see the discussions in \citetalias{2020Karunakaran} and \citealt{Read+2016}), and we therefore do not use it here.

Distinguishing the UDGs and LSB dwarfs based on their apparent properties beyond their central surface brightness is a difficult task (see Figures \ref{fig:optpropcomp} and \ref{fig:gasrichness-colour} and the discussion in $\S$\ref{subsec:udgdef}).\ An additional consideration of the morphological differences between our samples is to use quantitative non-parametric properties, such as the Gini coefficient \citep{2003Abraham,2004Lotz} or Concentration-Asymmetry-Smoothness (CAS) parameters \citep{2003Conselice}.\ These quantitative metrics describe different characteristics of a galaxy's stellar light distributions and we estimate these parameters using the \texttt{STATMORPH} Python package \citep{2019Rodriguez-Gomez} in a follow-up study (Karunakaran et al.\ in prep).\

\subsection{Formation and Evolution of UDGs}\label{subsec:formation}
One of the goals of this survey is to provide some additional constraints on the potential formation mechanisms of UDGs in low-density, field-like environments.\ We first discuss the bursty star formation mechanisms proposed by \citet{2017dicintio} using the NIHAO simulations before briefly discussing our sample in the context of other formation mechanisms.\ 

Within the bursty star-formation feedback model, bursts of episodic star formation in a UDG's evolution would push out and ``expand'' the stellar, gas, and, in turn, dark matter distributions, resulting in the extended sizes and low surface brightness expected for UDGs.\ A prediction from this model is a trend between gas richness ($\mhi/M_*$) and optical size ($R_{\mathrm{eff}}$) at a fixed stellar mass ($M_*$).\ We have previously investigated this trend with smaller samples of UDGs (e.g.\ \citealt{HCGUDGs}, \citetalias{2020Karunakaran}) and find, albeit tentative, evidence for a scaling between gas richness and size.\ 

We expanded our investigation of this trend using our full sample of \ion{H}{1} detections from the full survey data in Section \ref{subsec:detections} and Figure \ref{fig:gasrichness-size}.\ In contrast to the results from our pilot survey, Figure \ref{fig:gasrichness-size} shows that the UDGs and the LSB dwarfs in our sample have similar slopes in this plane within their uncertainties.\ Initially, this may suggest that this trend is not necessarily indicative of UDGs forming through a bursty star-formative feedback scenario.\ However, while it is clear that there is a large scatter in the best-fit slopes we measure (through multiple approaches, see Appendix \ref{sec:AltFit}), it is interesting that the median slopes are systematically higher for the UDGs and may further suggest a surface brightness dependence for the trends between $\mhi/M_*$ and $R_{\mathrm{eff}}$.\ Indeed, this also holds when we only consider the LSB dwarfs with large sizes (i.e.\ $R_{\mathrm{eff}}>1.5$ kpc).\ Furthermore, it is worth noting the relatively restrictive surface brightness criteria employed here.\ All things considered, our results imply that it is unclear that the trends seen in the $\mhi/M_*$-$R_{\mathrm{eff}}$ plane emerge as a result of the bursty star-formation feedback scenario, or simply as a selection effect.\

Beyond the predicted trend between $\mhi/M_*$ and $R_{\mathrm{eff}}$, the two subsets of outliers in our observed sample in the $\mhi/\LG$ -- $(g-r)$ plane appear to exhibit the properties one might expect from the cyclic nature of this bursty star formation scenario.\ The blue gas-poor systems are expected after a burst of star formation where gas is expelled, leaving a young (i.e.\ optically blue) and gas-deficient system.\ Further along in this cycle, the red gas-rich systems are expected after the stellar population has aged and gas has accreted into the system.\ The source of this gas could either be from the previously expelled, now-cooled gas or from the intergalactic medium.\ This leaves an older (i.e.\ optically red) and gas-rich system.\ While this is qualitatively consistent with the cyclic nature of this scenario, it is not necessarily unique to forming UDGs.\ 

The properties of our UDGs (and LSB dwarfs) are broadly consistent with predictions from various models.\ The low velocity widths of our sample combined with their large physical sizes suggest that they may be embedded in a high-spin halo \citep[e.g.][]{2016AmoriscoLoeb,2017Rong}.\ Although, as demonstrated by \citet[][]{2021Wright} the halo spin at early times is not necessarily comparable to its present-day spin.\ The relatively low SFRs and extended sizes of our sample imply that these systems have low SFR densities.\ \citet{2020ManceraPina} also see this for their sample of UDGs and suggest it may result from weak and inefficient internal feedback process (i.e.\ supernova feedback).\ The fact that these models make predictions that are also present for LSB dwarfs on the periphery of the UDG definition suggests that several formation mechanisms are likely at play.\ We expand this brief discussion to a more complete comparison between our sample of \ion{H}{1} detections and populations of UDGs and LSB dwarfs from the NIHAO and ROMULUS simulations suite in Motiwala et al.\ (in prep).\

\subsection{The effect of UDG definitions}\label{subsec:udgdef}
We briefly discuss how the definition of a UDG, specifically its surface brightness, can affect the results we have presented here.\ As shown in detail using dwarf galaxies from the ROMULUS simulations, several definitions have been used to select UDGs from LSB samples and can significantly affect the final selected subsets \citep[][]{2022VanNest}.\ 

We consider a broader surface brightness criterion using the $\langle\mu_{\mathrm{eff},g}\rangle$\footnote{We calculate this value using $\mathrm{m}_g$ and $r_{\mathrm{eff}}$ following \citet{2005Graham}: $\langle\mu_{\mathrm{eff},g}\rangle=\mathrm{m}_g + 2.5\mathrm{log}(2\pi r_{\mathrm{eff}}^2$).} to define UDGs from our \ion{H}{1} detection sample, akin to the definition used in some previous works \citep[e.g.][]{2017vanderBurg}.\ If we impose a limit of $\langle\mu_{\mathrm{eff},g}\rangle \geq\,24\,\mathrm{mag\,arcsec^{-2}}$ and keeping the same size limit of $R_{\mathrm{eff}}\geq 1.5$ kpc, then our sample sizes effectively flip to have 90 UDGs and 20 LSB dwarfs.\

\begin{figure*}[!ht]
\includegraphics[width=18cm]{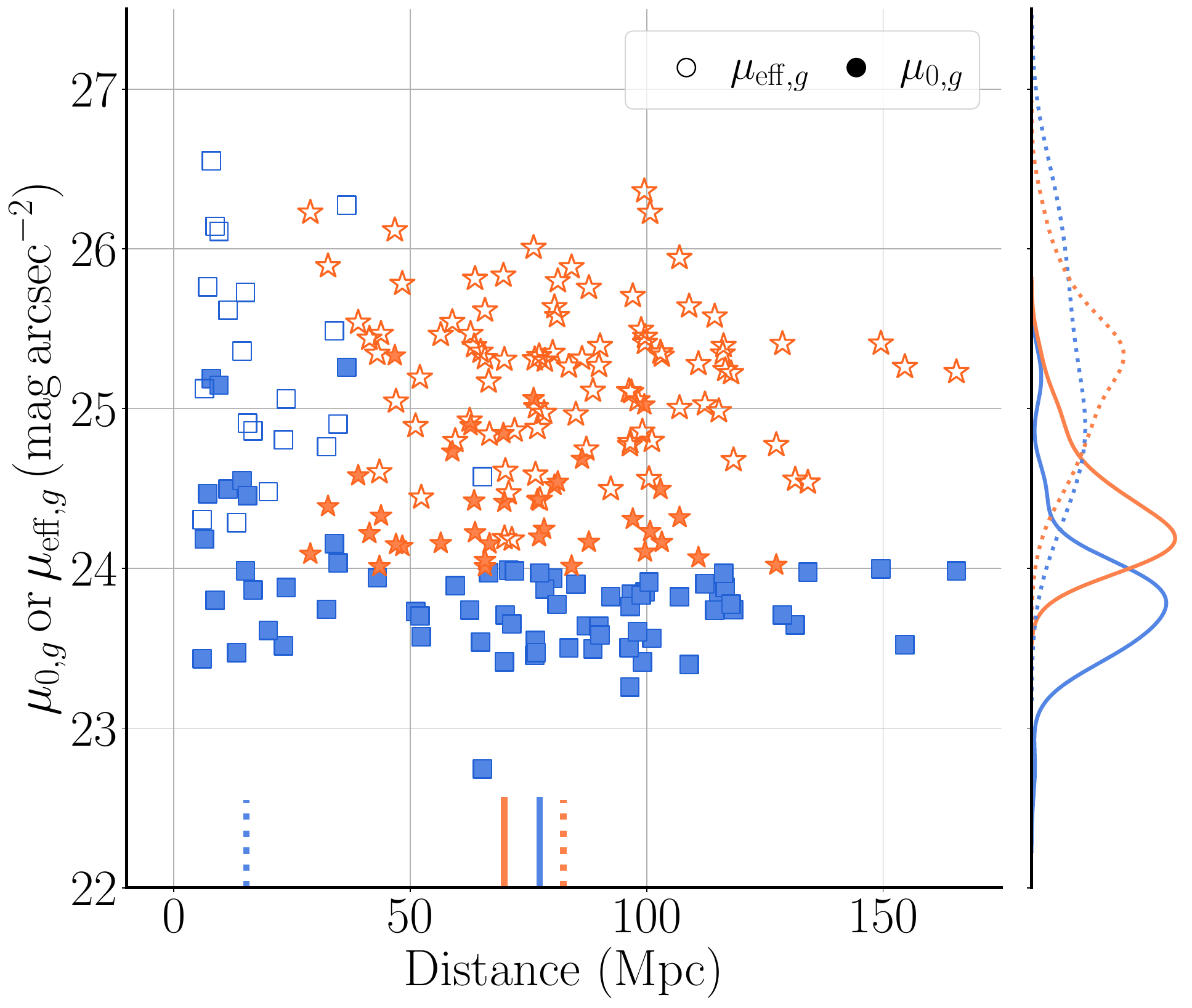}
\caption{$g-$band surface brightness as a function of distance for our UDGs (stars) and dwarfs (squares).\ The filled symbols show UDGs and dwarfs defined using $g-$band central surface brightness, $\mu_{0,g}$, while open symbols show those defined using their $g-$band effective surface brightness ($\langle\mu_{\mathrm{eff},g}\rangle$, see text).\ The marginalized surface brightness distributions are shown along the right-hand axis with the solid lines showing $\mu_{0,g}$ and dotted lines showing $\langle\mu_{\mathrm{eff},g}\rangle$.\ The vertical solid and dotted lines show the median distances for UDGs and dwarfs for the same surface brightness definitions.}
\label{fig:mudef}
\end{figure*}

The most notable shift in property between the samples as selected using $\mu_{0,g}$ and $\langle\mu_{\mathrm{eff},g}\rangle$ is their distance distribution.\ This result is unsurprising given the overlap in apparent magnitude and $r_{\mathrm{eff}}$ leading to an overlap in $\langle\mu_{\mathrm{eff},g}\rangle$.\ This leaves the physical effective radius, a distance-dependent quantity, as the primary selection cut.\ In Figure \ref{fig:mudef}, we show $\mu_{0,g}$ (solid symbols) and $\langle\mu_{\mathrm{eff},g}\rangle$ (open symbols) as a function of distance for our UDGs (orange stars) and LSB dwarfs (blue squares).\ The marginalized surface brightness distributions for each sample are placed along the right of the figure with solid lines for $\mu_{0,g}$ and dotted lines for $\langle\mu_{\mathrm{eff},g}\rangle$.\ As mentioned above, the clearer separation between UDG and LSB dwarf subsets can be seen in the $\langle\mu_{\mathrm{eff},g}\rangle$ sample.\ The vertical bars along the bottom of the figure show the median distances of each subset using each surface brightness definition: $\langle\mu_{\mathrm{eff},g}\rangle$ - $D_{UDG}=70$ Mpc, $D_{Dwarf}=77$ Mpc; $\mu_{0,g}$ - $D_{UDG}=82$ Mpc, $D_{Dwarf}=15$ Mpc.\

While the individual distances to these systems do not change, the separation in distances between UDG and LSB dwarf introduced by the alternative surface brightness definition propagates to all distance-dependent quantities.\ That is to say, UDGs and LSB dwarfs appear more distinct in various relations such as $\mhi-\mstar$, $\mhi/\mstar-R_{\mathrm{eff}}$, and SFR$-\mstar$.\ More relevant to some of the discussion concerning UDG formation mechanisms, we repeat our analysis in searching for a trend between $\mhi/\mstar$ as a function of $R_{\mathrm{eff}}$ using these new subsets.\ We find stronger evidence for this trend in our UDGs compared to our dwarfs: median slopes for most UDG stellar mass bins are consistently higher than the slopes for the dwarf bins that have median slopes near zero.\ This stark change in evidence for a predicted trend for a UDG formation mechanism is analogous to the results from \citet{2022VanNest}.\ However, the change here contrasts their results where we find that shifting from the ``traditional'' definition \citep[i.e.][]{2015vandokkum} to a broader one separates the UDG and dwarf populations, whereas \citet{2022VanNest} find that this change mixes the two populations, albeit for different physical properties.\ 

There are at least two takeaways from this exercise.\ Firstly, and perhaps obviously, using unified definitions to select UDG/LSB dwarf samples will remove any uncertainty when comparing one study to another.\ Secondly, it may be more informative and physically motivated to select objects as significant outliers from established scaling relations \citep[e.g.][]{lim2020,2023Li}.\

\section{Summary and Outlook} \label{sec:Conclusion}
We have presented the results from our GBT \ion{H}{1} follow-up survey of 378 optically-detected SMUDGes candidates.\ In total, we detect the \ion{H}{1} reservoirs of 110 targets and estimate their \ion{H}{1} properties (i.e.\ $\vsys$, $\wftyct$, and $\int S dv$) from the spectra through a combination of traditional \citep[e.g.][]{2005Springob} and more novel \citep[i.e.][]{2014Westmeier} methods.\ Crucially, we estimate the distances to our \ion{H}{1} detections from their $\vsys$ and use them to convert the angular sizes to physical ones to distinguish between UDGs and LSB dwarfs.\ Using these physical sizes, $R_{\mathrm{eff}}$, and the $g-$band central surface brightness, we confirm 37 targets as UDGs and 73 as low surface brightness dwarf galaxies.\ Additionally, we derive \ion{H}{1} masses, $\mhi$, from the spectra as well as stellar masses, $M_{*}$.\ To the best of our knowledge, this is the largest targeted \ion{H}{1} follow-up survey of optically selected UDG candidates to date.\

We presented our comparison of the derived \ion{H}{1} properties for our sample to that of the \ion{H}{1}-bearing Ultra Diffuse (HUD) sources \citep{2017Leisman,2019Janowiecki} finding commensurate results between the samples, albeit with ours having marginally lower gas richness.\ Similarly consistent with the HUDs sample, our \ion{H}{1} detections fall towards the faint and low mass end when compared to the broader ALFALFA catalog \citep{Haynes2018}.\ When comparing our \ion{H}{1} detections and non-detections, we find that, as expected, the former tend to have higher $\mhi/\LG$ while both our UDGs and LSB dwarfs have virtually identical median $\mhi/\LG$.\ 

We briefly presented the star-forming properties for our sample showing that our \ion{H}{1} detections follow the general trend of the broader gas-rich population (Figure \ref{fig:sfr_mstar}) but have marginally lower specific SFRs.\ Curiously, many of our candidates from Campaigns 4 and 5 selected to have signs of UV emission, and confirmed to have NUV and/or FUV emission, are not detected in our \ion{H}{1} observations.\ In terms of their environments, it is difficult to distinguish the \ion{H}{1} detections and non-detections in the sky (Figure \ref{fig:skydist}), however, the majority of our detections tend to reside in relatively isolated environments (i.e.\ no neighbors within $R_{proj}= 500\,$kpc or $\Delta \vsys = \pm 500\,\kms$) compared to our non-detections.\ 

The most significant distinguishing factor between the \ion{H}{1} detections and non-detections in our sample is their $g-r$ color (Figures \ref{fig:optpropcomp} and \ref{fig:gasrichness-colour}), especially when we consider only the subset of non-detections from Campaigns 1-3, where we did not impose the additional \textit{GALEX} UV emission criterion.\ Similarly, we find that qualitatively and quantitatively (Karunakaran et al.\ in prep) the \ion{H}{1} detections tend to have more irregular, clumpy morphologies compared to the smoother, more symmetric morphologies of the non-detections.\ Interestingly, we find that using an alternative, broader surface brightness definition, $\langle\mu_{\mathrm{eff},g}\rangle$, when selecting our UDGs, there is a distinct separation between UDGs and LSB dwarfs in their distance distributions.\

With a larger sample of \ion{H}{1}-confirmed UDGs and LSB dwarfs, we compared the properties of our sample to the predicted trends of some UDG formation mechanisms.\ We find that both our sample of UDGs and LSB dwarfs appear to have a positive correlation between their gas richness, $\mhi/M_{*}$, and size, $R_{\mathrm{eff}}$, at a given stellar mass.\ While this is commensurate with the predicted trend from the bursty star-formation feedback model from the NIHAO simulations, the fact that we also see this trend for our LSB dwarfs (within uncertainties) suggests that perhaps another mechanism is at play.\ Although we find stronger evidence for this trend using a broader surface brightness criterion for selecting UDGs, the validity is complicated by the trend's dependence on the selection criteria.\

Our survey has provided a significant sample of optically selected UDG candidates with an unprecedented set of \ion{H}{1} detections and predominantly stringent upper limits on our non-detections.\ We will use this sample to carry out a variety of additional studies on the properties of UDGs as a whole.\ Karunakaran et al.\ (in prep) will dive deeper into the star formation properties of our sample and present the trends (or lack thereof) on optical and \ion{H}{1} properties as well as their environments.\ In Motiwala et al.\ (in prep), we compare the \ion{H}{1} and star-forming properties of our sample to UDGs uniformly selected from the NIHAO and ROMULUS simulations, aiming to better constrain UDG formation mechanisms from these simulations.\ Already underway are multiple programs to obtain sufficiently resolved \ion{H}{1} imaging observations using radio interferometers.\ These resolved observations will enable detailed studies on the kinematics and morphologies of UDGs, significantly expanding the number of UDGs with resolved \ion{H}{1} imaging available in the literature to date.\ 

\floattable
\begin{deluxetable}{ccccCCCCCCCCCcCCc}[!ht]
\tablecaption{Target SMUDGes Candidate Properties \label{table:maintable}}
\rotate
\tabletypesize{\tiny}
\tablehead{
\colhead{Name} & \colhead{RA} & \colhead{Dec} & \colhead{Campaign} & \colhead{$m_g$} & \colhead{$\mu_{0,g}$} & \colhead{$g-r$} & \colhead{$r_{\mathrm{eff}}$} & \colhead{$b/a$} & \colhead{${\theta}$} & \colhead{$n$} & \colhead{Int.\ Time} & \colhead{${\sigma_{50}}$} & \colhead{HI}& \colhead{Distance} & \colhead{$\mathrm{log[\mhi/\LG]}$} & \colhead{Ref}\\
\colhead{} & \colhead{H:M:S} & \colhead{D:M:S} & \colhead{} & \colhead{(mag)} & \colhead{$(\mathrm{\frac{mag}{arcsec^{2}}})$} & \colhead{(mag)} & \colhead{(arcsec)} &\colhead{} & \colhead{(deg)} & \colhead{} & \colhead{(minutes)} & \colhead{(mJy)}  & \colhead{Det?} & \colhead{(Mpc)} & \colhead{$\mathrm{log[\msun/\lsun]}$} & \colhead{}\\
\colhead{(1)} & \colhead{(2)} & \colhead{(3)} & \colhead{(4)} & \colhead{(5)} & \colhead{(6)} & \colhead{(7)} & \colhead{(8)} & \colhead{(9)} & \colhead{(10)} & \colhead{(11)} & \colhead{(12)} & \colhead{(13)} & \colhead{(14)} & \colhead{(15)} & \colhead{(16)} & \colhead{(17)}
} 
\startdata
SMDG1103517+284120 & 11:03:51.7 & +28:41:20 & 1 & 17.80^{+0.15}_{-0.19} & 25.15^{+0.14}_{-0.07} & 0.31^{+0.26}_{-0.24} & 18.31^{+2.27}_{-0.42} & 0.85^{+0.02}_{-0.04} & -30^{+7}_{-6} & 0.81^{+0.09}_{-0.02} & 12 & 1.39 & Y & 9.5 & 0.51 & K24\\
SMDG1217378+283519 & 12:17:37.8 & +28:35:19 & 1 & 18.94^{+0.22}_{-0.06} & 24.93^{+0.19}_{-0.06} & 0.55^{+0.23}_{-0.23} & 10.40^{+0.42}_{-1.03} & 0.70^{+0.03}_{-0.03} & 47^{+3}_{-3} & 0.74^{+0.04}_{-0.09} & 30 & 0.58 &   &  & <0.09 & Z23 (K21)\\
SMDG1217443+332043 & 12:17:44.3 & +33:20:43 & 1 & 18.00^{+0.27}_{-0.13} & 25.48^{+0.18}_{-0.07} & 0.47^{+0.30}_{-0.30} & 17.51^{+1.50}_{-1.31} & 0.84^{+0.02}_{-0.04} & 78^{+6}_{-6} & 0.64^{+0.06}_{-0.05} & 30 & 0.69 &   &  & <-0.21 & Z23\\
SMDG1217451+281724 & 12:17:45.1 & +28:17:24 & 1 & 20.01^{+0.25}_{-0.07} & 25.66^{+0.24}_{-0.08} & 0.65^{+0.26}_{-0.23} & 7.36^{+0.38}_{-0.74} & 0.80^{+0.05}_{-0.07} & -1^{+9}_{-10} & 0.55^{+0.05}_{-0.10} & 144 & 0.38 &   &  & <0.33 & Z23\\
SMDG1220188+280131 & 12:20:18.8 & +28:01:31 & 1 & 18.47^{+0.23}_{-0.06} & 24.39^{+0.19}_{-0.04} & 0.35^{+0.24}_{-0.25} & 12.18^{+0.55}_{-1.30} & 0.65^{+0.02}_{-0.02} & -68^{+2}_{-2} & 0.95^{+0.04}_{-0.10} & 12 & 0.79 & Y & 32.6 & 0.58 & Z23\\
SMDG1220212+290831 & 12:20:21.2 & +29:08:31 & 1 & 19.46^{+0.19}_{-0.06} & 24.26^{+0.22}_{-0.03} & 0.60^{+0.20}_{-0.18} & 5.95^{+0.30}_{-0.40} & 0.92^{+0.03}_{-0.04} & -21^{+15}_{-16} & 0.90^{+0.05}_{-0.09} & 78 & 0.41 &   &  & <0.15 & Z23\\
SMDG1221086+292920 & 12:21:08.6 & +29:29:20 & 1 & 18.67^{+0.23}_{-0.07} & 25.20^{+0.17}_{-0.07} & 0.64^{+0.25}_{-0.23} & 14.17^{+0.52}_{-1.57} & 0.59^{+0.03}_{-0.02} & -3^{+2}_{-2} & 0.71^{+0.04}_{-0.08} & 24 & 0.73 &   &  & <0.08 & Z23 (K21)\\
SMDG1221235+303643 & 12:21:23.5 & +30:36:43 & 1 & 18.48^{+0.02}_{-0.31} & 25.47^{+0.13}_{-0.17} & 0.53^{+0.33}_{-0.31} & 19.42^{+4.47}_{0.61} & 0.63^{+0.02}_{-0.05} & 66^{+4}_{-3} & 0.89^{+0.20}_{0.00} & 36 & 0.84 &   &  & <0.06 & Z23\\
SMDG1221401+284346 & 12:21:40.1 & +28:43:46 & 1 & 19.25^{+0.18}_{-0.08} & 24.98^{+0.18}_{-0.07} & 0.62^{+0.20}_{-0.18} & 9.36^{+0.37}_{-0.72} & 0.60^{+0.03}_{-0.02} & 76^{+2}_{-3} & 0.65^{+0.05}_{-0.07} & 54 & 0.53 &   &  & <0.18 & Z23\\
SMDG1221497+283111 & 12:21:49.7 & +28:31:11 & 1 & 18.85^{+0.33}_{-0.08} & 25.75^{+0.22}_{-0.07} & 0.55^{+0.34}_{-0.32} & 16.15^{+0.83}_{-2.33} & 0.63^{+0.03}_{-0.04} & 9^{+3}_{-3} & 0.69^{+0.06}_{-0.10} & 30 & 0.85 &   &  & <0.22 & Z23\\
\enddata

\tablecomments{The first 10 rows of this table are shown here.\ The full table is available online in machine-readable format.\ \\ \\ col.(1): Adopted SMUDGes UDG candidate name.\ cols.(2) and (3): J2000 position of optical centroid, which corresponds to our GBT LOS.\ col.(4): Campaign number.\ cols.(5) - (7): $g-$band apparent magnitude, central surface brightness, and $g-r$ colors.\ col.\ (8)-(11): Best-fitting effective radius, axial ratio, position angle, and $\mathrm{S\acute{e}rsic}$ index of the UDG candidate with corresponding uncertainties derived as in \citetalias{2023Zaritsky}.\ col.(12): Total effective GBT integration time, including the ON+OFF positions and subtracting any time lost due to RFI.\ col.(13): Representative RMS noise of the spectrum at a velocity resolution of $\Delta V = 50 \, \kms$.\ col.(14): \ion{H}{1} detection?.\ col.(15): Derived or adopted distance.\ Distances derived from this work will have a ``Y'' in col.(14), otherwise distance reference in parentheses in col.(17).\ col.(16): Logarithm of $\mhi/\LG$ for \ion{H}{1} detections, see col.(14), otherwise upper limit on logarithm of $\mhilim/\LG$ and marked with '$<$'.\ col.(17): Reference from which SMUDGes candidate is selected, alternative distance references in parentheses.\ K24 = This work, Z23 = \citetalias{2023Zaritsky}; alternative references for distances K12 = \citet{2012Kochanek}, K14 = \citet{2014KimExVirgo}, vD15 = \citet{2015vandokkumb}, G18 = \citet{2018Gu}, RL18 = \citet{2018RuizLara}, K21 = \citet{2021Kadowaki}, C22 = \citet{2022Carlsten}, S24 = \citet{2024Shen}.\ }
\end{deluxetable}

\setlength{\tabcolsep}{3pt}
\begin{deluxetable*}{cCCCCCCCCCC}[htb!]
\caption{Properties of UDGs with \ion{H}{1} detections}
\label{table:detectiontable}
\tablehead{
\colhead{Name} & \colhead{$\Delta V$} & \colhead{$\sigma_{\Delta V}$} & \colhead{$V_{sys}$} & \colhead{$\wftyct$} & \colhead{$S_{HI}$} & \colhead{$D_{HI}$} & \colhead{log($\mhi$)} & \colhead{log($M_{*}$)} & \colhead{log$\mbary$} & \colhead{$R_{\mathrm{eff}}$} \\
\colhead{} & \colhead{($\kms$)} & \colhead{(mJy)} & \colhead{($\kms$)} & \colhead{($\kms$)} & \colhead{(Jy$\,\kms$)} & \colhead{(Mpc)} & \colhead{(log[$\msun$])} & \colhead{(log[$\msun$])} & \colhead{(log[$\msun$])} & \colhead{(kpc)} \\
\colhead{(1)} & \colhead{(2)} & \colhead{(3)} & \colhead{(4)} & \colhead{(5)} & \colhead{(6)} & \colhead{(7)} & \colhead{(8)} & \colhead{(9)} & \colhead{(10)} & \colhead{(11)}
}
\startdata
SMDG0015544+280910 & 15 & 0.50 & 8918\pm1 & 31\pm14 & 0.26\pm0.00 & 127 & 8.99\pm0.03 & 8.42^{+0.20}_{-0.21} & 9.10^{+0.05}_{-0.05} & 3.53^{+0.19}_{-0.14}\\
SMDG0018470+302845 & 25 & 0.46 & 5484\pm5 & 42\pm17 & 0.12\pm0.03 & 78 & 8.24\pm0.12 & 8.63^{+0.20}_{-0.21} & 8.78^{+0.15}_{-0.15} & 2.74^{+0.19}_{-0.18}\\
SMDG0025100+310314 & 15 & 0.63 & 4687\pm28 & 24\pm5 & 0.39\pm0.15 & 67 & 8.61\pm0.18 & 8.18^{+0.20}_{-0.21} & 8.75^{+0.14}_{-0.14} & 1.97^{+0.16}_{-0.15}\\
SMDG0033164+283310 & 10 & 0.66 & 4878\pm2 & 60\pm1 & 0.27\pm0.04 & 70 & 8.50\pm0.09 & 7.97^{+0.20}_{-0.21} & 8.61^{+0.08}_{-0.09} & 2.80^{+0.22}_{-0.21}\\
SMDG0050362+322104 & 10 & 1.97 & 4926\pm4 & 18\pm4 & 0.41\pm0.03 & 66 & 8.62\pm0.07 & 8.80^{+0.20}_{-0.22} & 9.02^{+0.12}_{-0.13} & 4.74^{+0.40}_{-0.49}\\
SMDG0126070-020411 & 10 & 1.00 & 2020\pm1 & 67\pm2 & 1.34\pm0.27 & 29 & 8.42\pm0.17 & 7.78^{+0.20}_{-0.22} & 8.51^{+0.15}_{-0.15} & 1.74^{+0.31}_{-0.38}\\
SMDG0214097+283647 & 10 & 0.87 & 2896\pm1 & 40\pm1 & 0.31\pm0.03 & 41 & 8.10\pm0.11 & 7.92^{+0.20}_{-0.21} & 8.32^{+0.11}_{-0.11} & 1.76^{+0.22}_{-0.23}\\
SMDG0728567+293658 & 15 & 0.61 & 4894\pm3 & 58\pm4 & 0.32\pm0.00 & 70 & 8.56\pm0.06 & 8.28^{+0.20}_{-0.21} & 8.75^{+0.08}_{-0.08} & 3.14^{+0.24}_{-0.23}\\
SMDG0740191+232853 & 10 & 0.73 & 5409\pm2 & 63\pm1 & 0.57\pm0.01 & 77 & 8.90\pm0.06 & 8.07^{+0.20}_{-0.21} & 8.96^{+0.06}_{-0.06} & 2.62^{+0.20}_{-0.20}\\
SMDG0833162+280607 & 15 & 0.48 & 6144\pm1 & 31\pm6 & 0.13\pm0.00 & 88 & 8.36\pm0.05 & 8.17^{+0.20}_{-0.23} & 8.58^{+0.08}_{-0.09} & 3.11^{+0.22}_{-0.38}\\
\enddata
\tablecomments{The first 10 rows of this table are shown here.\ The full table is available online in machine-readable format.\ col.(2): Velocity resolution of spectrum used to compute \ion{H}{1} properties.\ col.(3): RMS noise of spectrum at $\Delta V$ in col.(2).\ col.(4): Heliocentric systemic velocity.\ col.(5): Velocity width of the \ion{H}{1} detection, corrected for cosmological redshift and instrumental broadening, as well as ISM turbulence.\ col.(6): Integrated \ion{H}{1} flux.\ col.(7): Distance estimated using the Hubble-Lema\^{i}tre Law, $\vsys$ and $\mathrm{H}_0 = 70 \, \kms\,\mathrm{Mpc}^{-1}$.\ We adopt distance uncertainties of 5 Mpc.\ col.(8): Logarithm of \ion{H}{1} mass calculated from Eq.\ref{eqn:himass} using $S_{HI}$ in col.(6) and $D_{HI}$ in col.(7).\ col.(9): Logarithm of stellar mass calculated using $m_g$ and $g-r$ from Table \ref{table:maintable}, $D_{HI}$ in col.(7), and the corresponding relation from \citet{2017zhang}.\ col.(10): Logarithm of baryonic mass, $1.33\mhi + M_{*}$.\ col.(11): Effective radius in physical units using $r_{\mathrm{eff}}$ from Table \ref{table:maintable} and $D_{HI}$ in col.(7).\ }
\end{deluxetable*}

\begin{deluxetable*}{cCCCCCCCCCC}[htb!]
\caption{\ion{H}{1} Properties of LSB dwarfs}
\label{table:dwarftable}
\tablehead{
\colhead{Name} & \colhead{$\Delta V$} & \colhead{$\sigma_{\Delta V}$} & \colhead{$V_{sys}$} & \colhead{$\wftyct$} & \colhead{$S_{HI}$} & \colhead{$D_{HI}$} & \colhead{log($\mhi$)} & \colhead{log($M_{*}$)} & \colhead{log$\mbary$} & \colhead{$R_{\mathrm{eff}}$}\\
\colhead{} & \colhead{($\kms$)} & \colhead{(mJy)} & \colhead{($\kms$)} & \colhead{($\kms$)} & \colhead{(Jy$\,\kms$)} & \colhead{(Mpc)} & \colhead{(log[$\msun$])} & \colhead{(log[$\msun$])} & \colhead{(log[$\msun$])} & \colhead{(kpc)} \\
\colhead{(1)} & \colhead{(2)} & \colhead{(3)} & \colhead{(4)} & \colhead{(5)} & \colhead{(6)} & \colhead{(7)} & \colhead{(8)} & \colhead{(9)} & \colhead{(10)} & \colhead{(11)}
}
\startdata
SMDG0006565+244829 & 15 & 0.59 & 4380\pm1 & 45\pm1 & 0.38\pm0.01 & 62.6 & 8.54\pm0.07 & 7.99^{+0.20}_{-0.21} & 8.65^{+0.07}_{-0.07} & 1.69^{+0.15}_{-0.14}\\
SMDG0012551+290947 & 50 & 0.41 & 6774\pm35 & 36\pm7 & 0.76\pm0.32 & 96.5 & 9.22\pm0.19 & 8.65^{+0.20}_{-0.21} & 9.33^{+0.16}_{-0.16} & 3.57^{+0.21}_{-0.19}\\
SMDG0016164+250812 & 10 & 0.61 & 5952\pm1 & 20\pm3 & 0.19\pm0.03 & 85.0 & 8.50\pm0.08 & 8.08^{+0.20}_{-0.20} & 8.64^{+0.08}_{-0.08} & 2.08^{+0.13}_{-0.13}\\
SMDG0031012+294037 & 15 & 0.75 & 5359\pm6 & 60\pm10 & 0.51\pm0.03 & 76.6 & 8.85\pm0.06 & 7.36^{+0.20}_{-0.20} & 8.86^{+0.06}_{-0.06} & 1.89^{+0.14}_{-0.13}\\
SMDG0045409+305950 & 10 & 0.84 & 5045\pm1 & 41\pm5 & 0.34\pm0.02 & 72.1 & 8.62\pm0.06 & 8.08^{+0.20}_{-0.21} & 8.73^{+0.07}_{-0.07} & 2.19^{+0.16}_{-0.15}\\
SMDG0049279+282117 & 10 & 0.85 & 4994\pm21 & 67\pm8 & 0.65\pm0.16 & 71.5 & 8.89\pm0.12 & 8.11^{+0.20}_{-0.21} & 8.96^{+0.11}_{-0.11} & 1.70^{+0.13}_{-0.12}\\
SMDG0121260+310806 & 10 & 1.61 & 4909\pm1 & 31\pm1 & 0.70\pm0.10 & 70.1 & 8.91\pm0.09 & 8.30^{+0.20}_{-0.21} & 9.01^{+0.08}_{-0.08} & 2.60^{+0.20}_{-0.19}\\
SMDG0137165+323415 & 10 & 1.90 & 6308\pm3 & 18\pm2 & 0.65\pm0.07 & 90.1 & 9.09\pm0.07 & 8.53^{+0.20}_{-0.21} & 9.20^{+0.07}_{-0.07} & 4.32^{+0.27}_{-0.40}\\
SMDG0143289+322415 & 10 & 1.52 & 8163\pm1 & 36\pm1 & 0.74\pm0.07 & 116.6 & 9.38\pm0.05 & 8.76^{+0.20}_{-0.21} & 9.47^{+0.06}_{-0.06} & 5.09^{+0.27}_{-0.26}\\
SMDG0146339+331915 & 10 & 1.26 & 5417\pm1 & 67\pm2 & 1.36\pm0.09 & 77.4 & 9.28\pm0.06 & 8.31^{+0.20}_{-0.21} & 9.33^{+0.06}_{-0.06} & 2.27^{+0.16}_{-0.15}\\
\enddata
\centering
\tablecomments{The first 10 rows of this table are shown here.\ The full table is available online in machine-readable format.\ All parameters have the same definitions as in Table \ref{table:detectiontable}.\ }
\end{deluxetable*}
\section*{Acknowledgments}
AK acknowledges support from the Natural Sciences and Engineering Research Council of Canada (NSERC), the University of Toronto Arts \& Science Postdoctoral Fellowship program, and the Dunlap Institute.\ 

KS acknowledges support from NSERC.

DZ acknowledges support from NSF AST-1713841 and AST-2006785 and NASA ADAP 80NSSC23K0471.

This research has made use of the NASA/IPAC Extragalactic Database (NED), which is operated by the Jet Propulsion Laboratory, California Institute of Technology, under contract with the National Aeronautics and Space Administration.

The Legacy Surveys consist of three individual and complementary projects: the Dark Energy Camera Legacy Survey (DECaLS; Proposal ID \#2014B-0404; PIs: David Schlegel and Arjun Dey), the Beijing-Arizona Sky Survey (BASS; NOAO Prop. ID \#2015A-0801; PIs: Zhou Xu and Xiaohui Fan), and the Mayall z-band Legacy Survey (MzLS; Prop. ID \#2016A-0453; PI: Arjun Dey). DECaLS, BASS and MzLS together include data obtained, respectively, at the Blanco telescope, Cerro Tololo Inter-American Observatory, NSF’s NOIRLab; the Bok telescope, Steward Observatory, University of Arizona; and the Mayall telescope, Kitt Peak National Observatory, NOIRLab. Pipeline processing and analyses of the data were supported by NOIRLab and the Lawrence Berkeley National Laboratory (LBNL). The Legacy Surveys project is honored to be permitted to conduct astronomical research on Iolkam Du’ag (Kitt Peak), a mountain with particular significance to the Tohono O’odham Nation.

See \url{https://www.legacysurvey.org/acknowledgment/#scientific-publication-acknowledgment} for the full Legacy Surveys acknowledgment.\

The \textit{GALEX} data presented in this paper were obtained from the Mikulski Archive for Space Telescopes (MAST) at the Space Telescope Science Institute. The specific observations analyzed can be accessed via \dataset[https://doi.org/10.17909/5q50-gy07]{https://doi.org/10.17909/5q50-gy07}. STScI is operated by the Association of Universities for Research in Astronomy, Inc., under NASA contract NAS5–26555. Support to MAST for these data is provided by the NASA Office of Space Science via grant NAG5–7584 and by other grants and contracts.
\vspace{5mm}
\facilities{GBT (VEGAS)}

\software{astropy \citep{astropy:2018}, astroquery \citep{astroquery}, asymmetric\_uncertainty \citep{2022gobat}, dustmaps \citep{dustmaps}, GBTIDL \citep{GBTIDL}, linmix \citep{2007Kelly}, matplotlib \citep{matplotlib}, numpy \citep{numpy}, pandas \citep{pandas}, photutils \citep{photutils}, seaborn \citep{seaborn}, scipy \citep{scipy}, uncertainties \citep{uncertainties} }

\appendix

\section{Comparison of Derived \ion{H}{1} Properties with Literature Values}\label{sec:a100comp}
\begin{figure}[!ht]
\includegraphics[width=15cm]{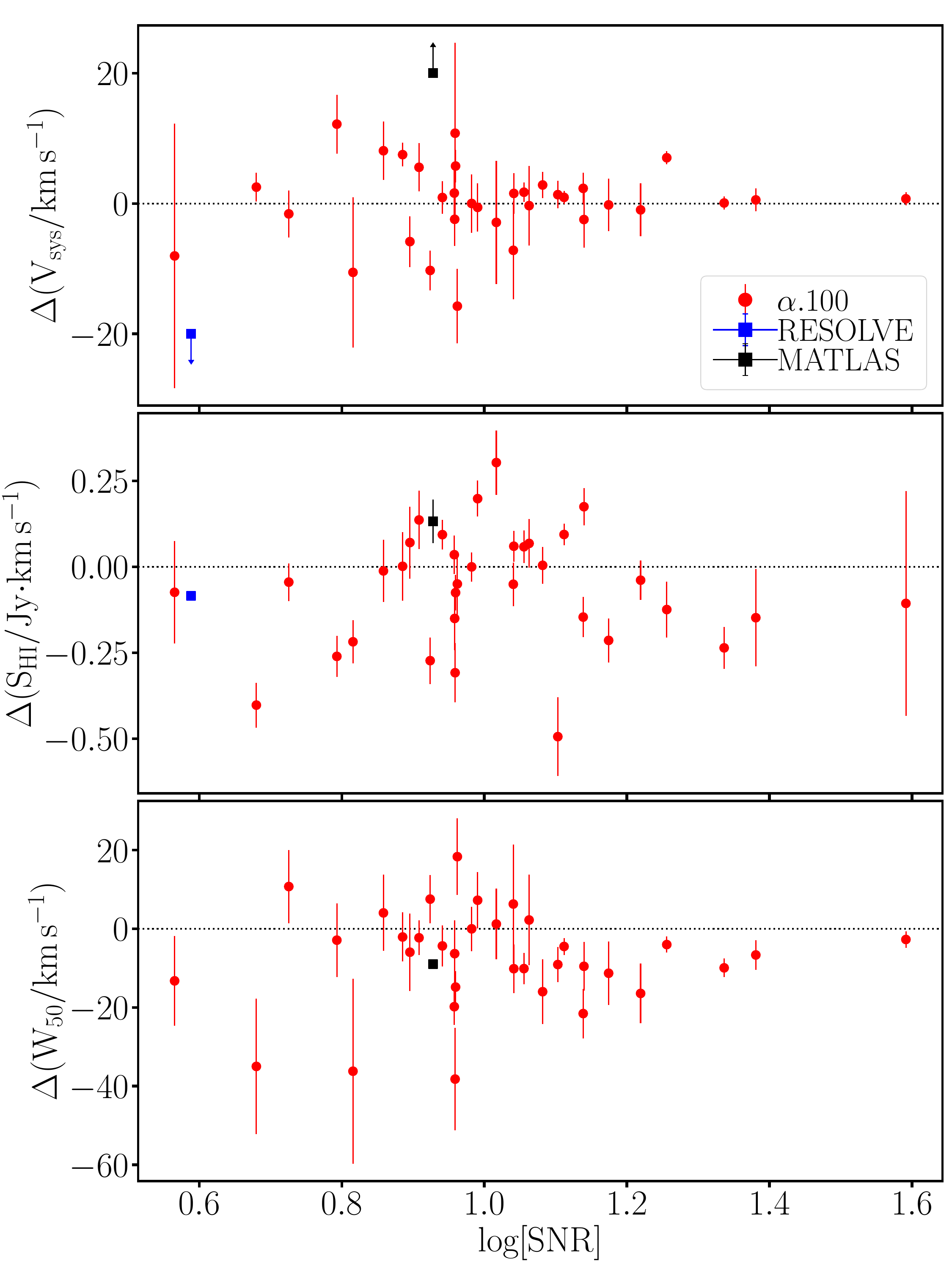}
\caption{Comparisons of the difference between the systemic velocities ($\Delta\vsys$, top panel), \ion{H}{1} fluxes ($\Delta{S_{HI}}$, middle panel), and velocity widths ($\Delta\wfty$, bottom panel) as a function of the logarithm of the signal to noise ratio (SNR) in the GBT data between the overlapping galaxies in our sample and literature \ion{H}{1} observations.\ The sources overlapping the ALFALFA $\alpha.100$ catalog are shown as red inverted triangles, the source from the RESOLVE survey is shown as a blue square, and the source from the MATLAS survey is shown as a black square.\ In the top panel, we show the RESOLVE and MATLAS sources with downward and upward-facing arrows to indicate their larger offsets from the GBT value.\ We note that the RESOLVE source does not provide uncertainties on $S_{HI}$ or quote a $\wfty$ value and the MATLAS source does not have uncertainties on $\wfty$.\ }
\label{fig:smudges_a100_props}
\end{figure}

Here we compare the derived \ion{H}{1} properties of the subset of galaxies that overlap between our observations and those from the literature.\ In Figure \ref{fig:smudges_a100_props}, we show the difference between the systemic velocity ($\Delta\vsys$, top panel), \ion{H}{1} flux ($\Delta{S_{HI}}$, middle panel), and velocity width ($\Delta\wfty$, bottom panel) as a function of the logarithm of the signal to noise ratio (SNR) as measured in our analysis.\ The error bars show the uncertainties in our measured quantities and the literature values added in quadrature.\ Many of the deltas in these derived properties fall within the uncertainties of the measured values and most fall within the overall scatter.\ However, the derived recessional velocities from our data are more significantly offset for the two sources from RESOLVE \citep[blue square with downward arrow][]{2023Hutchens} and MATLAS \citep[black square with upward arrow][]{2022Poulain}.\ While the offset for the RESOLVE source is more modest, $V_{sys,GBT} - V_{RESOLVE} \sim-78\kms$, the difference for the MATLAS source is substantial, $V_{sys,GBT} - V_{MATLAS} \sim166\kms$.\ It is unclear why such difference is present in just $\vsys$ values for these two sources, particularly that from MATLAS, while the offsets in $S_{HI}$ and $\wfty$ are well within the estimated uncertainties.

\section{Alternative Approaches to Fitting Gas Richness Size}\label{sec:AltFit}
\begin{figure*}[!ht]
\includegraphics[width=8cm]{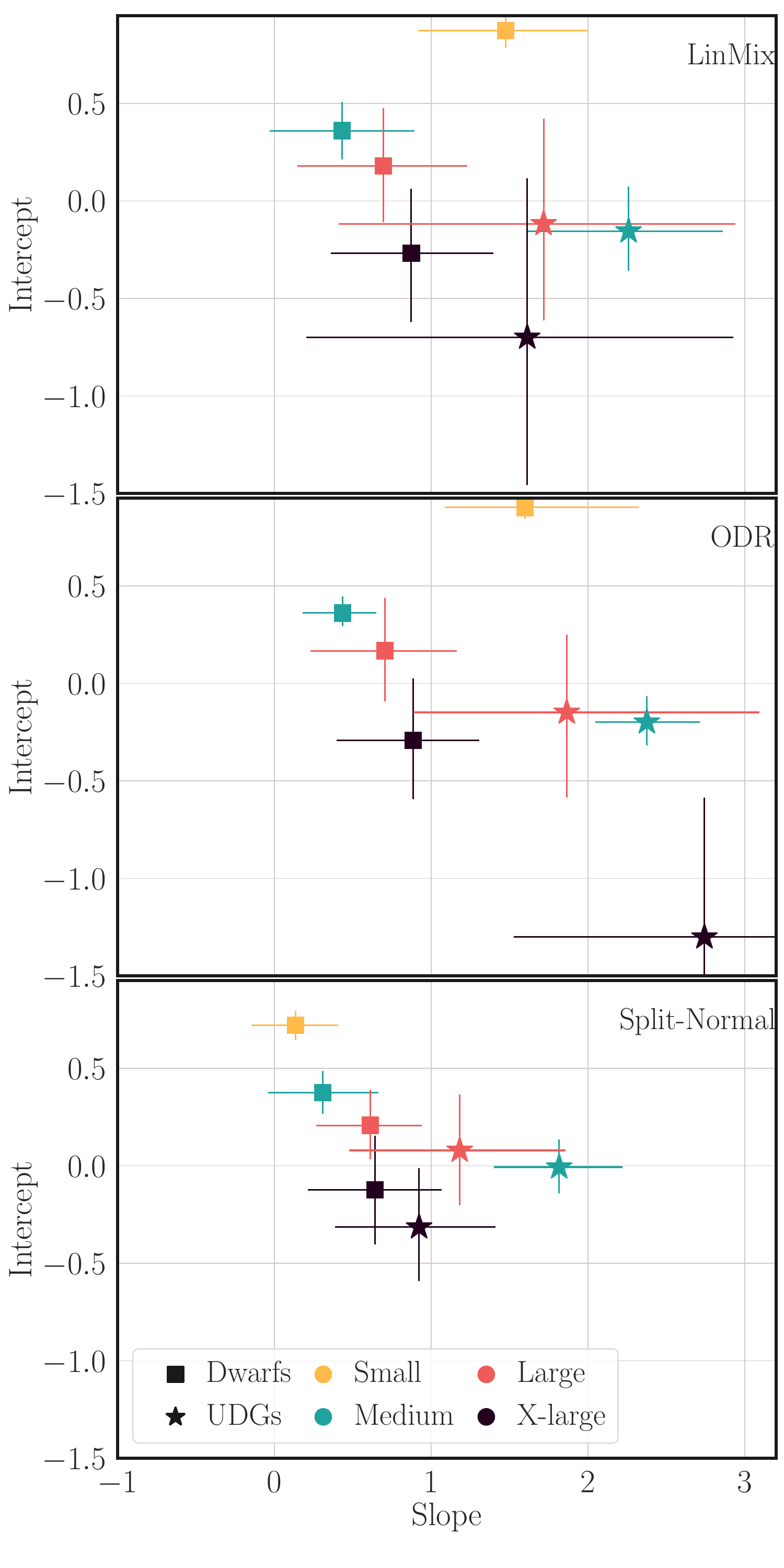}
\includegraphics[width=8cm]{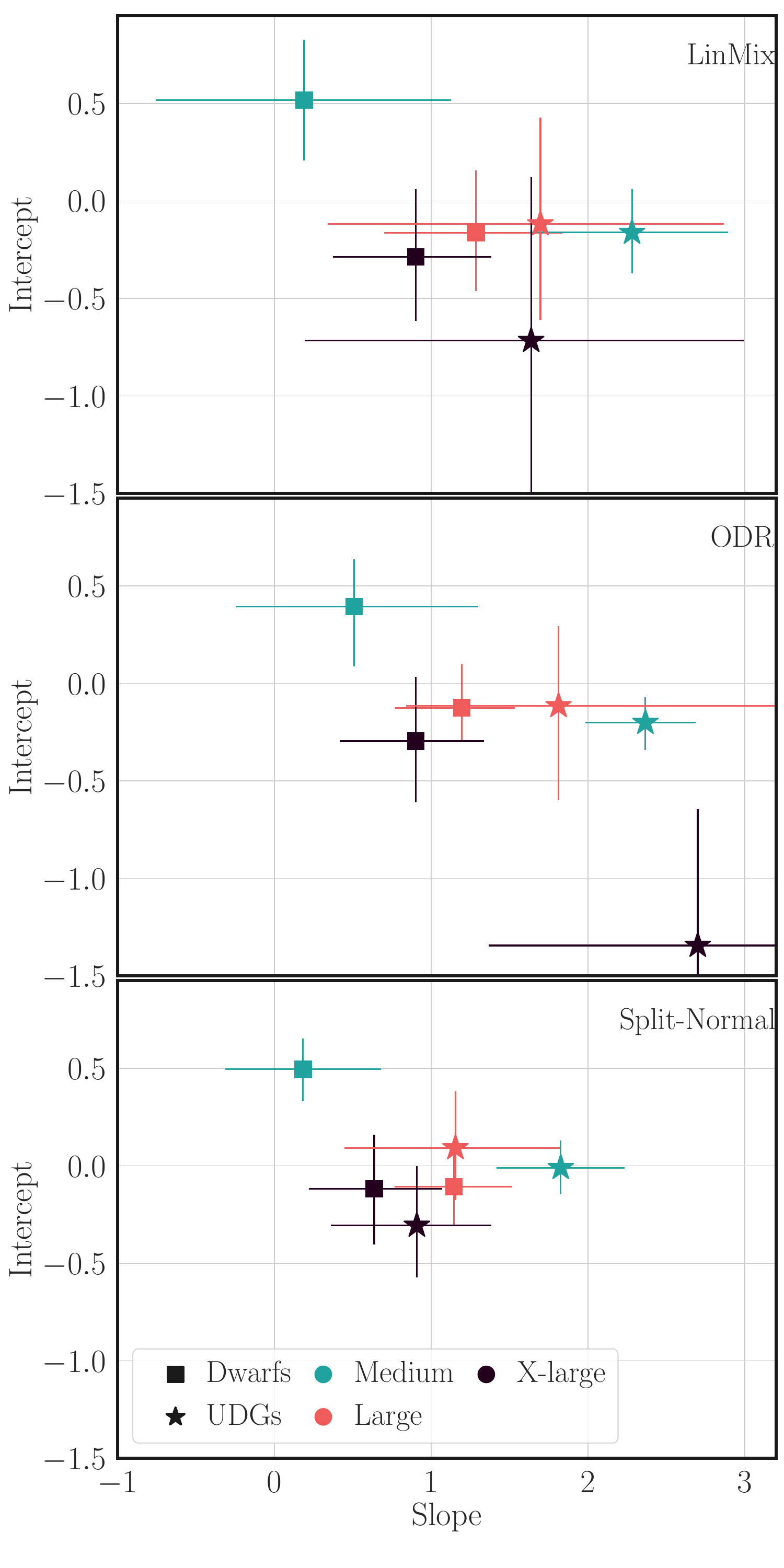}
\caption{Comparison of the best-fit slope and intercept in the $\mathrm{log}[\mhi/M_{*}]$-$\mathrm{log}[R_{\mathrm{eff}}]$ plane.\ Each row shows the best-fit parameters for each of our fitting procedures: top - \texttt{LinMix}, middle - ODR, and bottom - Split-Normal.\ The symbol color represents their stellar mass bin as in Figure \ref{fig:gasrichness-size} and the symbol type represents LSB dwarfs (squares) and UDGs (stars).\ The left panels include all of our \ion{H}{1} detections, whereas the right panels only include dwarf galaxies with $R_{\mathrm{eff}}>1.5$ kpc.}
\label{fig:fitcomp_approaches}
\end{figure*}

We perform two alternative fitting procedures to the \texttt{LinMix} methodology described in Section~\ref{sec:results} to obtain the best-fit relation to our UDGs and LSB dwarfs in the $\mathrm{log}[\mhi/M_{*}]$-$\mathrm{log}[R_{\mathrm{eff}}]$ plane shown in Figure \ref{fig:gasrichness-size}.\ First, we use Orthogonal Distance Regression (ODR\footnote{https://docs.scipy.org/doc/scipy/reference/odr.html}) to obtain the best-fit relation for each subset.\ We note that, as with our fiducial approach using \texttt{LinMix}, we assume symmetric uncertainties for our ODR procedure again selecting whichever is larger of the lower and upper bounds.\ Furthermore, we perform 1000 bootstrap resampling iterations of the ODR and take the median and interquartile range (IQR) as the best-fit values and scatter for the slope and intercept of the relation.\ In our second alternative approach, we perform a Monte Carlo sampling from a Split-Normal distribution where each ``piece'' of the distribution has a width defined by the upper and lower uncertainty for a given value.\ The properties of each source are sampled from this distribution and a simple linear regression is performed 10000 times.\ As in our ODR method, we take the median and IQR for the slope and intercept as our best-fit parameters.\ In the left panels of Figure \ref{fig:fitcomp_approaches}, we show the slope (x-axis) and intercept (y-axis) for each of our methods: top - \texttt{LinMix}, middle - ODR, and bottom - Split-Normal.\ The color of the symbols represents the stellar mass bins as in Figure \ref{fig:gasrichness-size}.\ The results for the UDGs and LSB dwarfs are represented by the star and square symbols, respectively.\ In the right panels, we show the same results now with dwarf galaxies selected to have $R_{\mathrm{eff}}>1.5$ kpc as a comparison between large LSB dwarf galaxies and UDGs.\

The results across all three approaches suggest the same broad conclusions: i) UDGs have larger median slopes than LSB dwarfs and ii) there are no significant differences for most of the UDGs and LSB dwarfs in each stellar mass bin within their IQR.\ There are two, however, two exceptions of note.\ First, in the Medium stellar mass bin (green symbols), the slope and intercept are statistically different across all three methods.\ Second, in the X-large stellar mass bin for the ODR approach the UDGs and LSB dwarfs are also statistically different as their IQR do not overlap.\ While we believe the first of these exceptions due to their consistency across all fitting approaches, we find it difficult to accept the second, especially given the large scatter in both slope and intercept for the UDGs.\

\bibliography{references}{}
\bibliographystyle{aasjournal}

\end{document}